\documentclass[a4paper,11pt]{article}
\pdfoutput=1

\usepackage[T1]{fontenc} 
\usepackage{tikz}
\usepackage{tikz-cd}
\usepackage{
graphicx,
hyperref,
amsmath,
amssymb,
charter,
xcolor,
ifluatex,
longtable,
booktabs,
multirow,
bbold,cancel}

\usepackage{
graphicx,
hyperref,
amsmath,
amssymb,
charter,
xcolor,
ifluatex,
longtable,
booktabs,
multirow,
bbold}
        
\usepackage{colortbl}

\definecolor{Gray}{gray}{0.95}
\definecolor{RGray}{gray}{0.85}
\definecolor{CGray}{gray}{0.92}

\usepackage{multicol}
\definecolor{tit}{rgb}{0.1,0.2,0.4}
\definecolor{blus}{cmyk}{1,1,0,0.6}
\definecolor{verde}{cmyk}{0.92,0,0.59,0.25}

\usepackage{tikz}
\usetikzlibrary{matrix}

\usepackage{devanagari}
\usepackage{tipa}
\usepackage{amsmath,amssymb,amsfonts, bm}
\usepackage{dsfont}
\usepackage{epsfig}
\usepackage{graphicx}
\usepackage{slashed}
\usepackage{color}

\usepackage{caption}
\usepackage{subcaption}
\usepackage{slashed}

\usepackage{commath}
\usepackage{calc}

\newcommand{\be}{\begin{equation}}
\newcommand{\ee}{\end{equation}}

\newcommand{\bea}{\begin{eqnarray}}
\newcommand{\eea}{\end{eqnarray}}

\newcommand{\bfig}{\begin{figure}}
\newcommand{\efig}{\end{figure}}

\usepackage{slashed} 

\newcommand{\xdownarrow}[1]{%
  {\left\updownarrow\vbox to #1{}\right.\kern-\nulldelimiterspace}
}

\usepackage{commath}
\usepackage{calc}

\newcommand\scalemath[2]{\scalebox{#1}{\mbox{\ensuremath{\displaystyle #2}}}}

\newcommand{\D}{{\cal D}}
\newcommand{\U}{{\cal U}}

\newcommand{\N}{{\cal N}}
\newcommand{\M}{{\cal M}}

\makeatletter
\newcommand*{\rom}[1]{\expandafter\@slowromancap\romannumeral #1@}
\makeatother

\usepackage{color,hyperref}
\definecolor{darkblue}{rgb}{0.0,0.0,0.3}
\hypersetup{colorlinks,breaklinks,
            linkcolor=darkblue,urlcolor=darkblue,
            anchorcolor=darkblue,citecolor=darkblue}

\def\1by3{\ensuremath{\frac{1}{3}}}
\def\4by3{\ensuremath{\frac{4}{3}}}
\def\2by3{\ensuremath{\frac{2}{3}}}

\providecommand*\url[1]{\href{#1}{#1}}
\renewcommand*\url[1]{\href{#1}{\texttt{#1}}}

\begin{document} 

\allowdisplaybreaks
\vspace*{-2.5cm}
\begin{flushright}
{\small
IIT-BHU
}
\end{flushright}

\vspace{2cm}

\begin{center}
{\LARGE \bf \color{tit}
 Origin of the  VEVs hierarchy}\\[1cm]

{\large\bf Gauhar Abbas$^{a}$\footnote{email: gauhar.phy@iitbhu.ac.in}   }  
\\[7mm]
{\it $^a$ } {\em Department of Physics, Indian Institute of Technology (BHU), Varanasi 221005, India}\\[3mm]

\vspace{1cm}
{\large\bf\color{blus} Abstract}
\begin{quote}
We present an origin of the VEVs hierarchy in a non-minimal technicolour framework which is capable of explaining the flavour spectrum of the standard model along with neutrino masses and mixing, and simultaneously satisfying crucial experimental bounds.  The technicolour scale in this framework can be lower such that  a standard model-like Higgs boson emerges from within the model.  We also derive lower bound on the mass of the vector technicolour state using the latest experimental bound on the $S$-parameter.
\end{quote}

\thispagestyle{empty}
\end{center}

\begin{quote}
{\large\noindent\color{blus} 
}

\end{quote}

\newpage
\setcounter{footnote}{0}

\section{Introduction}
\label{intro}
 The fermionic mass hierarchy and mixing in the Standard Model (SM), often referred as to the flavour problem, is an intriguing puzzle\cite{Abbas:2017vws}-\cite{King:2013hoa}.  This problem evidently arises because of the hierarchical Yukawa couplings which appear in the Higgs-fermion Lagrangian of the SM.  For more references on this problem, see ref.\cite{Abbas:2017vws}.

One possible way to explain the masses and mixing of the SM is through hierarchical vacuum expectation values (VEVs) which introduce new energy scales which can distinguish the SM fermions among and within the  generations\cite{Abbas:2017vws}.   This can be done via appropriate non-renormalizable operators responsible for providing masses to fermions. For instance the simplest choice is  dimension-5 operators of the form,
\be
y~  \bar{\psi}_L \varphi \psi_R \dfrac{F_i}{\Lambda},
\ee
where $y$ is dimensionless coupling, $\psi_L$ is the SM left-handed fermionic doublet, $\psi_R$ denotes the SM right-handed singlet fermions, $\varphi$ is the Higgs doublet,  $F_i$ represents  quantum fields, and $\Lambda$ characterizes scales where the operator is renormalized.

The  different energy scales which can distinguish the fermion among and within the generations  are created   through the  hierarchical  VEVs  of  fields $F_i$ which are actively present in the effective theory.  Here, there is only one hidden scale $\Lambda$ which is integrated out.  

Before we discuss further, let us investigate how many energy scales are required to explain the fermionic mass pattern, for instance the masses of six quarks.  In the SM, the masses of the first family fermions are much smaller than that of the second family fermions, and masses of the second family fermions are much smaller than that of the third family fermions.  This is known as the mass hierarchy among the families.  This can be explained by introducing exactly three new interactions at three different energy scales where one scale may correspond to only one family.  

We now look into the mass hierarchy within the families.  This is large within the second and third families requiring two energy scales to differentiate the masses within these families.  However, in the first family of quarks, the mass of the $d$-quark has the same order of magnitude as that of the $u$-quark.  This is the origin of the strong isospin symmetry of the proton and neutron.  Therefore, it is not required to introduce any new energy scale within the first family.  Thus we conclude that for explaining masses of the three families of quarks through new energy scales, atleast five energy scales are required.  

However, the mass pattern within the first family is subtle, and presumably more challenging to predict than that of within the second and third families.  The subtlety lies in the fact that the mass of the $d$-quark is slightly bigger than that of the $u$-quark.  This is starkly different from second and third families where the masses of $up$-type quarks are much larger than the $down$-type quarks. This is also the origin of the strong isospin breaking.  To find an explanation to this intriguing difference is a problem itself  within the flavour problem.

In the hierarchical VEVs  model (HVM)\cite{Abbas:2017vws}, the dimension-5 operators are obtained  by forbidding the Yukawa Lagrangian with the help of  three discrete $\mathcal{Z}_2$ symmetries and   introducing six gauge-singlet scalar fields.  The masses and mixing of the SM fermions are explained through the hierarchy of the  six VEVs of the  gauge-singlet scalar fields.  There is no hierarchy in the effective Yukawa couplings now.  However, the hierarchy of Yukawa couplings is now converted into  the hierarchy of  six VEVs.   Finding an origin of this VEVs hierarchy is the main  objective of this work.

We shall show that an origin of the HVM scenario can  arise in a Technicolour (TC) framework through multi-fermion chiral condensates.  TC framework provides a natural origin of the spontaneous symmetry breaking of the SM, and may address several issues beyond the SM\cite{Weinberg:1975gm,Susskind:1978ms}.  Flavour structure of the SM can also be addressed  in an   extended TC (ETC) framework \cite{ETC}.  However, it is well known that  ETC models contain large flavour changing neutral currents (FCNC) involving fermions\cite{ETC}.  

This problem can be cured by modifying the TC dynamics where the anomalous mass dimension of composite operator $\bar{T}T$ of the TC field $T$ becomes large\cite{large1,large2,large3}.  There exits  two kind of such solutions.   The first solution is with the slowly running coupling constant, known as walking gauge theories\cite{walking1}-\cite{walking7}.  In the second solution, we have a gauge theory with strong four-fermion interaction\cite{strong-4F-1}-\cite{DN2}.   For a review  of TC theories, see refs. \cite{far,Kaul:1981uk,mira,Lane:1993wz,Kitazawa:1994pd,Chivukula:1998if,Chivukula:2000mb,Hill:2002ap,Andersen:2011yj}.

This work  will be presented along the following line.  In the section \ref{HVM} we present a  hierarchical VEVs Model (HVM)\cite{Abbas:2017vws} and discuss  the  neutrino masses and oscillations.  Moreover, numerical fitting of fermion masses and mixing angles, the scalar potential and an ultra-violet completion (UV) of the HVM are also presented in the same section.  A TC origin of the HVM in a non-minimal conventional TC framework is discussed in section \ref{TC_model}.  We summarize in section \ref{sum}.  
\section{Hierarchical VEVs Model}
\label{HVM}
The essential idea of the HVM is discussed in ref.\cite{Abbas:2017vws}\footnote{An alternative  HVM model is discussed in the appendix.}.  In this work, we present a new manifestation of this idea.  In the HVM, we extend the gauge symmetry of the SM  by adding the three Abelian discrete symmetries  $\mathcal{Z}_2$, $\mathcal{Z}_2^\prime$ an $\mathcal{Z}_2^{\prime \prime} $.  Moreover, we add six gauge singlet scalar fields  $\chi _i $ ($i=1-6$), having zero hypercharge, to the SM.   The  fermionic and the singlet scalar  fields  $\chi _i $  transform under the symmetries $\mathcal{Z}_2$,  $\mathcal{Z}_2^\prime$ and $\mathcal{Z}_2^{\prime \prime}$ as given in  Table \ref{tab1}. 

The transformations of the scalar fields  $\chi _i $  under the $SU(3)_c \otimes SU(2)_L \otimes U(1)_Y$ symmetry of the SM are given as,
\begin{eqnarray}
\chi _i :(1,1,0),
 \end{eqnarray} 
where $i=1-6$.

The Yukawa Lagrangian is forbidden now, and the  masses of the charged fermions  are recovered by dimension-5 operators  written as,
\bea
\label{mass2}
{\mathcal{L}}_{mass} &=& \dfrac{1}{\Lambda_{\rm F} }\Bigl[  y_{ij}^u  \bar{\psi}_L^{q,a}  \tilde{\varphi} \psi_R^{u, b}   \chi _i +     
   y_{ij}^d  \bar{\psi}_L^{q,a}   \varphi \psi_R^{d, b}  \chi _{i+3}   +   y_{ij}^\ell  \bar{\psi}_L^{\ell, a}   \varphi \psi_R^{\ell, b}  \chi _{i+3} \Bigr]  
+  {\rm H.c.},
\eea
where $\Lambda_{\rm F}$ is the scale where these operators are renormalized.  The  $\psi_L^{q,\ell, i}$ represents the leptonic and quark  doublets of the SM, and $a, b=1,2,3$ are family indices, $q$ is for the quarks doublet and $\ell$ denotes the leptonic doublet, and $\psi_R^u =  u_R, c_R, t_R, ~ \psi_R^d = d_R, s_R, b_R, \psi_R^\ell = e_R, \mu_R, \tau_R $.

 The masses of charged fermions  are recovered when fields $\chi _i$ acquire VEVs in such a way that  $ \langle \chi _4 \rangle > \langle \chi _1 \rangle $, $ \langle \chi _2 \rangle >> \langle \chi _5 \rangle $, $ \langle \chi _3 \rangle >> \langle \chi _6 \rangle $, $ \langle \chi _{3} \rangle >> \langle \chi _{2} \rangle >> \langle \chi _{1} \rangle $, and  $ \langle \chi _6 \rangle >> \langle \chi _5 \rangle >> \langle \chi _4 \rangle $.  
\begin{table}[h]
\begin{center}
\begin{tabular}{|c|c|c|c|}
  \hline
  Fields             &        $\mathcal{Z}_2$                    & $\mathcal{Z}_2^\prime$   & $\mathcal{Z}_2^{\prime \prime}$      \\
  \hline
  $u_{R}, c_{R}, t_{R}, \nu_{e_R}, \nu_{\mu_R}, \nu_{\tau_R}$                 &   -  &     -    & -                                \\
  $\chi _1$                        & -  &      -  &+                                                 \\
   $\chi _2$                        & -  &      +  &-                                                 \\
    $\chi _3$                        & +  &      -  &-                                                 \\
   $d_{R} $, $ s_{R}$, $b_{R}$,  $e_R$, $ \mu_R$,  $\tau_R$  &   +  &     -    & -                        \\
   $\chi _4$                        & +  &      -  &+                                                 \\
   $\chi _5$                        & +  &      +  &-                                                 \\
    $\chi _6$                        & -  &      -  &-                                                 \\
    $ \psi_L^1 $                        & +  &      +  &-                                                 \\
   $ \psi_L^2 $                         & +  &      -  &+                                                 \\
   $ \psi_L^3 $                         & -  &      +  &+                                                 \\
   $ \varphi $                         & +  &      +  &+                                                 \\
    $\chi_7 $                        & +  &      +  &+                                               \\
  \hline
     \end{tabular}
\end{center}
\caption{The charges of left- and right-handed fermions  and gauge singlet scalar fields under $\mathcal{Z}_2$, $\mathcal{Z}_2^\prime$ and $\mathcal{Z}_2^{\prime \prime}$ symmetries.}
 \label{tab1}
\end{table} 

The mass matrices of up, down quarks and leptons approximately are,
\begin{align}
\label{mUD}
\M_\U & =   \dfrac{ v }{\sqrt{2}}  \begin{pmatrix}
  y_{11}^u \epsilon_1 &  y_{12}^u \epsilon_1 &   y_{13}^u \epsilon_1 \\
  y_{21}^u \epsilon_2 &   y_{22}^u \epsilon_2 &   y_{23}^u \epsilon_2\\
   y_{31}^u \epsilon_3 &     y_{32}^u \epsilon_3  &   y_{33}^u \epsilon_3\\
\end{pmatrix} ,  \nonumber
\M_\D = \dfrac{ v }{\sqrt{2}}  \begin{pmatrix}
  y_{11}^d \epsilon_4 &    y_{12}^d \epsilon_4 &  y_{13}^d \epsilon_4 \\
  y_{21}^d \epsilon_5 &     y_{22}^d \epsilon_5 &   y_{23}^d \epsilon_5\\
    y_{31}^d \epsilon_6 &     y_{32}^d \epsilon_6  &   y_{33}^d \epsilon_6\\
\end{pmatrix}, 
\M_\ell =\dfrac{ v }{\sqrt{2}}  \begin{pmatrix}
  y_{11}^\ell \epsilon_4 &    y_{12}^\ell \epsilon_4 &   y_{13}^\ell \epsilon_4 \\
 y_{21}^\ell \epsilon_5 &    y_{22}^\ell \epsilon_5 &   y_{23}^\ell \epsilon_5\\
    y_{31}^\ell \epsilon_6 &     y_{32}^\ell \epsilon_6  &   y_{33}^\ell \epsilon_6 \\
\end{pmatrix},
\end{align} 
where $\epsilon_i = \dfrac{\langle \chi _{i} \rangle }{\Lambda_{\rm F}}$.   

The masses of charged fermions can approximately be written as \cite{Rasin:1998je},
\begin{eqnarray}
\label{mass1}
m_t  &\approx& \ |y^u_{33}| \epsilon_3 v/\sqrt{2}, \nonumber \\
m_c  &\approx& \   |y^u_{22} - \dfrac{y_{23}^u  y_{32}^u}{y_{33}^u}| \epsilon_2 v /\sqrt{2} ,\nonumber \\
m_u  &\approx&  |y_{11}^u - {y_{12}^u y_{21}^u \over  |y^u_{22} - \dfrac{y_{23}^u  y_{32}^u}{y_{33}^u}| }   -
{{y_{13}^u (y_{31}^u y_{22}^u - y_{21}^u y_{32}^u )-y_{31}^u  y_{12}^u  y_{23}^u } \over 
{ |y^u_{22} - \dfrac{y_{23}^u  y_{32}^u}{y_{33}^u}|  |y^u_{33}|}}   |\,  \epsilon_1 v /\sqrt{2},\nonumber \\
m_b  &\approx& \ |y^d_{33}| \epsilon_6 v/\sqrt{2}, \nonumber \\
m_s  &\approx& \   |y^d_{22} - \dfrac{y_{23}^d  y_{32}^d}{y_{33}^d}| \epsilon_5 v /\sqrt{2} ,\nonumber \\
m_d  &\approx&  |y_{11}^d - {y_{12}^d y_{21}^d \over  |y^d_{22} - \dfrac{y_{23}^d  y_{32}^d}{y_{33}^d}| }   -
{{y_{13}^d (y_{31}^d y_{22}^d - y_{21}^d y_{32}^d )-y_{31}^d  y_{12}^d  y_{23}^d } \over 
{ |y^d_{22} - \dfrac{y_{23}^d  y_{32}^d}{y_{33}^d}|  |y^d_{33}|}}   |\,  \epsilon_4 v /\sqrt{2},\nonumber \\
m_\tau  &\approx& \ |y^\ell_{33}| \epsilon_6 v/\sqrt{2}, \nonumber \\
m_\mu  &\approx& \   |y^\ell_{22} - \dfrac{y_{23}^\ell  y_{32}^\ell}{y_{33}^\ell}| \epsilon_5 v /\sqrt{2} ,\nonumber \\
m_e  &\approx&  |y_{11}^\ell - {y_{12}^\ell y_{21}^\ell \over  |y^\ell_{22} - \dfrac{y_{23}^\ell  y_{32}^\ell}{y_{33}^\ell}| }   -
{{y_{13}^\ell (y_{31}^\ell y_{22}^\ell - y_{21}^\ell y_{32}^\ell )-y_{31}^\ell  y_{12}^\ell  y_{23}^\ell } \over 
{ |y^\ell_{22} - \dfrac{y_{23}^\ell  y_{32}^\ell}{y_{33}^\ell}|  |y^\ell_{33}|}}   |\,  \epsilon_4 v /\sqrt{2}.
\end {eqnarray}
The quark mixing angles  approximately can be written using the equation (10) of reference \cite{Rasin:1998je},
\begin{eqnarray}
\sin \theta_{12}  \simeq |V_{us}| &\simeq& \left|{y_{12}^d \epsilon_4 \over y_{22}^d \epsilon_5}   -{y_{12}^u \epsilon_1 \over y_{22}^u \epsilon_2}  \right| , 
\sin \theta_{23}  \simeq |V_{cb}| \simeq  \left|{y_{23}^d \epsilon_5  \over y_{33}^d \epsilon_6 }  -{y_{23}^u \epsilon_2 \over y_{33}^u \epsilon_3 }  \right| ,\nonumber \\
\sin \theta_{13}  \simeq |V_{ub}| &\simeq& \left|{y_{13}^d  \epsilon_4  \over y_{33}^d  \epsilon_6 }  -{y_{12}^u y_{23}^d  \epsilon_1 \epsilon_5 \over y_{22}^u y_{33}^d  \epsilon_2  \epsilon_6} 
- {y_{13}^u \epsilon_1 \over y_{33}^u \epsilon_3} \right|.
\end{eqnarray}   
\subsection{Neutrino masses and oscillation parameters}
\label{neutrino}
Explaining neutrino masses and oscillation is one of the most challenging problems, and needs extension of the HVM by an additional gauge singlet scalar field  $\chi_7$ whose charges under  $\mathcal{Z}_2$,  $\mathcal{Z}_2^\prime$   and  $\mathcal{Z}_2^{\prime \prime}$ discrete symmetries are shown in table \ref{tab1}.

The Lagrangian for  Dirac masses for neutrinos  can be written as,
\bea
{\mathcal{L}}_{ \M_{\D} } &=&  \dfrac{1}{\Lambda_{\rm F}}     \left(   y_{ij}^\nu  \bar{\psi}_L^{\ell,i}  \tilde{\varphi_2} \psi_R^{\nu, j}   \chi _{1,2,3}      \right)
+  {\rm H.c.}, 
\eea
where  $\psi_R^\nu = \nu_{e_R}, \nu_{\mu_R}, \nu_{\tau_R} $.  

The Dirac mass matrix  is now  written as,
\begin{equation}
\label{mN}
\begin{array}{ll}
\M_\D = \dfrac{ v }{\sqrt{2}} \left( \begin{array}{ccc}
  y_{11}^\nu \epsilon_1  &    y_{12}^\nu \epsilon_1 &   y_{13}^\nu  \epsilon_1 \\
 y_{21}^\nu \epsilon_2 &    y_{22}^\nu \epsilon_2 &  y_{23}^\nu  \epsilon_2 \\
   y_{31}^\nu \epsilon_3 &   y_{32}^\nu \epsilon_3 &   y_{33}^\nu \epsilon_3
\end{array} \right).
\end{array}
\end{equation}
The  Majorana mass Lagrangian for right-handed neutrinos  reads,
\bea
\mathcal{L}_{ \M_{R}}  = \chi_7 \bar{\nu^c}_{i} \nu_{j}. 
\eea
The masses of neutrinos  now can be determined using type-\rom{1} seesaw mechanism\cite{seesaw}  providing  following mass matrix of the light neutrinos,
\begin{eqnarray}
{ \M}~ =~ -  \M_{\D}  \M_{R}^{-1}  \M_{\D}^T,
\end{eqnarray}
where $ \M_{\D} << \M_{R}$, and the Majorana mass terms for the left-handed neutrinos are assumed to be much smaller.   

The light neutrino masses   approximately are \cite{Rasin:1998je},
\begin{eqnarray}
m_3  &\approx& \ |y^\nu_{33}| \epsilon_3 \epsilon, \nonumber \\
m_2  &\approx& \   |y^\nu_{22} - \dfrac{y_{23}^\nu  y_{32}^\nu}{y_{33}^\nu}| \epsilon_2 \epsilon ,\nonumber \\
m_1  &\approx&  |y_{11}^\nu - {y_{12}^\nu y_{21}^\nu \over  |y^\nu_{22} - \dfrac{y_{23}^\nu  y_{32}^\nu}{y_{33}^\nu}| }   -
{{y_{13}^\nu (y_{31}^\nu y_{22}^\nu - y_{21}^\nu y_{32}^\nu )-y_{31}^\nu  y_{12}^\nu  y_{23}^\nu } \over 
{ |y^\nu_{22} - \dfrac{y_{23}^\nu  y_{32}^\nu}{y_{33}^\nu}|  |y^\nu_{33}|}}   |\,  \epsilon_1 \epsilon,
\end {eqnarray}
where $\epsilon = \frac{v} {\sqrt{2} M_R}$, and $M_R = \langle \chi_7 \rangle$ is the Majorana mass scale.

The leptonic mixing angles are approximately  found to be\cite{Rasin:1998je},
\begin{eqnarray}
\sin \theta_{12}^\nu  &\simeq& \left|{y_{12}^\ell \epsilon_4 \over y_{22}^\ell \epsilon_5}   -{y_{12}^\nu \epsilon_1 \over y_{22}^\nu \epsilon_2}  \right| , 
\sin \theta_{23}^\nu  \simeq  \left|{y_{23}^\ell \epsilon_5  \over y_{33}^\ell \epsilon_6 }  -{y_{23}^\nu \epsilon_2 \over y_{33}^\nu \epsilon_3 }  \right| ,\nonumber \\
\sin \theta_{13}^\nu  &\simeq& \left|{y_{13}^\ell  \epsilon_4  \over y_{33}^\ell  \epsilon_6 }  -{y_{12}^\nu y_{23}^\ell  \epsilon_1 \epsilon_5 \over y_{22}^\nu y_{33}^\ell  \epsilon_2  \epsilon_6} 
- {y_{13}^\nu \epsilon_1 \over y_{33}^\nu \epsilon_3} \right|.
\end{eqnarray}
We note that  the neutrino mixing angles turn out to be similar to those of the quark sector.  It  may be understood from the unification perspective. For instance, it is already known that there could be a scenario where unification of the neutrino and quark mixing is possible\cite{Mohapatra:2003tw}.   In such a scenario, the large neutrino mixing angles are the result of of renormalization group evolution \cite{AbdusSalam:2019hov}-\cite{Abbas:2013uqh}.
\subsection{Benchmark points for fermionic masses, quark-mixing and neutrino oscillation parameters }
\label{fit_eff}
In  the case of the  HVM, the need is to explain the whole  fermionic mass and mixing spectrum including neutrino oscillation parameters using just six hierarchical VEVs and the Majorana mass scale.    We particularly note that  the weak phases of the CKM matrix place extremely strong constraints on  models trying to explain origin of the flavour in the SM.  Since any theory  beyond  the SM has to explain neutrino masses and mixing, this makes survivability of any particular model of flavour miserable.  

The charged fermion masses  are reproduced using the following values of the fermion masses at 1 TeV\cite{Xing:2007fb},
\begin{eqnarray}
\label{xing2007}
\{m_t, m_c, m_u\} &\simeq& \{150.7 \pm 3.4,~ 0.532^{+0.074}_{-0.073},~ (1.10^{+0.43}_{-0.37}) \times 10^{-3}\}~{\rm GeV}, \nonumber \\
\{m_b, m_s, m_d\} &\simeq& \{2.43\pm 0.08,~ 4.7^{+1.4}_{-1.3} \times 10^{-2},~ 2.50^{+1.08}_{-1.03} \times 10^{-3}\}~{\rm GeV},
\nonumber \\
\{m_\tau, m_\mu, m_e\} &\simeq& \{1.78\pm 0.2,~ 0.105^{+9.4 \times 10^{-9}}_{-9.3 \times 10^{-9}},~ 4.96\pm 0.00000043 \times 10^{-4}\}~{\rm GeV}.
\end{eqnarray}
The magnitudes and phases  of the Cabibbo–Kobayashi–Maskawa (CKM) mixing elements are reproduced using following  available data\cite{Tanabashi:2018oca},
\bea
|V_{ud}| &=& 0.97446 \pm 0.0001, |V_{us}| = 0.22452 \pm 0.0004,  |V_{cb}| = 0.04214 \pm 0.00075, \\ \nonumber |V_{ub}| &=& 0.00365 \pm 0.00012, 
\sin 2 \beta = 0.691 \pm 0.017, ~ \alpha = (84.5^{+5.9}_{-5.2})^\circ,~  \gamma = (73.5^{+4.2}_{-5.1})^\circ
\eea

The neutrino oscillation parameters are given as \cite{deSalas:2017kay},
\bea
\Delta m_{21}^2 &=& (7.55^{+0.59}_{-0.5}) \times 10^{-5} {\rm eV}^2, |\Delta m_{31}^2| = (2.50\pm 0.09) \times 10^{-3} \rm{eV}^2,  \\ \nonumber
\sin^2 \theta_{12}^\nu &=&  (3.20^{+0.59}_{-0.47}) \times 10^{-1},
 \sin^2 \theta_{23}^\nu =  (5.47^{+0.52}_{-1.02}) \times 10^{-1},  \sin^2 \theta_{13}^\nu =  (2.160^{+0.25}_{-0.20}) \times 10^{-2},
\eea
where range of errors is $3 \sigma$.

The fermion masses and mixing are fitted by defining 
 \begin{align}
 \chi^2 &= \dfrac{(m_q - m_q^{\rm{model}} )^2}{\sigma_{m_q}^2}+  \dfrac{(m_\ell - m_\ell^{\rm{model}} )^2}{\sigma_{m_\ell}^2}  + \dfrac{(\sin \theta_{ij} -\sin \theta_{ij}^{\rm{model}} )^2}{\sigma_{\sin \theta_{ij}}^2}     + \dfrac{(\sin 2 \beta  -\sin 2 \beta^{\rm{model}} )^2}{\sigma_{\sin2\beta}^2} \\ \nonumber  
 &+ \dfrac{( \alpha    - \alpha^{\rm{model}} )^2}{\sigma_{\alpha}^2}  + \dfrac{( \gamma    - \gamma^{\rm{model}} )^2}{\sigma_{\gamma}^2} +  \dfrac{(\Delta m_{21}^2 - \Delta m_{21}^{2 ~ \rm{model}} ) }{\sigma_{\Delta m_{21}^2}^2} + \dfrac{(\Delta m_{31}^2 - \Delta m_{31}^{2 ~ \rm{model}} ) }{\sigma_{\Delta m_{31}^2}^2}  \\ \nonumber  
 &+ \dfrac{(\sin \theta_{ij}^\nu -\sin \theta_{ij}^{\nu ~\rm{model}} )^2}{\sigma_{\sin \theta_{ij}^\nu}} 
 \end{align} 
where $q=u,d,c,s,t,b$, $\ell=e,\mu,\tau$ and $i,j=1,2,3$.  

The weak phases of the CKM matrix in the standard choice are given by,
\begin{eqnarray}
\beta^{\text{model}} =\text{arg} \left(- \dfrac{V_{cd} V_{cb}^*}{V_{td} V_{tb}^*}\right),~\alpha^{\text{model}} =\text{arg} \left(- \dfrac{V_{td} V_{tb}^*}{V_{ud} V_{ub}^*}\right),~\gamma^{\text{model}} =\text{arg} \left(- \dfrac{V_{ud} V_{ub}^*}{V_{cd} V_{cb}^*}\right).
\end{eqnarray}
The dimensionless coefficients $y_{ij}^{u,d,\ell}= |y_{ij}^{u,d,\ell}| e^{i \phi_{ij}^{q,\ell}}$  are scanned with $|y_{ij}^{u,d,\ell}| \in [0.5,4\pi]$ and $ \phi_{ij}^{q,\ell} \in [0,2\pi]$, and corresponding coefficients for leptonic sector $y_{ij}^{\nu}= |y_{ij}^{\nu}| e^{i \phi_{ij}^{\nu}}$  are scanned with $|y_{ij}^{\nu}| \in [0.5, 4\pi]$ and $ \phi_{ij}^{\nu} \in [0,2\pi]$.

The result of fitting is,
\begin{align}
\{|y_{11}^u|, |y_{12}^u|, |y_{13}^u|,|y_{21}^u|, |y_{22}^u|, |y_{23}^u|,|y_{31}^u|, |y_{32}^u|, |y_{33}^u| \}  
&=  \{0.55, 0.67,
0.54, 1.48, 1.09, 1.1, 1.42, 0.54, 0.97\},  \nonumber \\
 \{\phi_{11}^u, \phi_{12}^u,  \phi_{13}^u,\phi_{21}^u,\phi_{22}^u,\phi_{23}^u,\phi_{31}^u, \phi_{32}^u,\phi_{33}^u \} 
 &=  \{ 
3.39, 1.05, 0.72, 1.41, 3.71, 3.56, 3.47, 0.98, 5.05\}, \nonumber \\
\{|y_{11}^d|, |y_{12}^d|, |y_{13}^d|, |y_{21}^d|, |y_{22}^d|, |y_{23}^d|, |y_{31}^d|, |y_{32}^d|, |y_{33}^d|\}  
& =  \{1.62, 2.17, 2.22, 2.16, 0.5, 1.17, 1.47, 0.52, 1.34\}, \nonumber \\
\{\phi_{11}^d, \phi_{12}^d,  \phi_{13}^d,  \phi_{21}^d,\phi_{22}^d,  \phi_{23}^d,\phi_{31}^d \phi_{32}^d,\phi_{33}^d\} & =  \{ 4.72, 5.91, 4.37, 0.52, 2.49, 5.70, 1.93, 1.63, 3.87    \}, \nonumber \\
\{|y_{11}^\ell|, |y_{12}^\ell|, |y_{13}^\ell|, |y_{21}^\ell|, |y_{22}^\ell|, |y_{23}^\ell|, |y_{31}^\ell|, |y_{32}^\ell|, |y_{33}^\ell|\} & =  \{  4.27,  12.57, 
    12.57,  12.47,  1.33, 
    10.63,   0.95,   0.5,   0.5 \}, \nonumber \\
\{\phi_{11}^\ell, \phi_{12}^\ell,  \phi_{13}^\ell,  \phi_{21}^\ell,\phi_{22}^\ell,\phi_{23}^\ell,\phi_{31}^\ell,\phi_{32}^\ell,\phi_{33}^\ell \} & =  \{  5.08,  5.31, 0.77,  0.64,  2.51,  1.45, 
   5.28,  1.02,  0.76  \}, \nonumber \\
 \{|y_{11}^\nu|, |y_{12}^\nu|, |y_{13}^\nu|, |y_{21}^\nu|, |y_{22}^\nu|, |y_{23}^\nu|, |y_{31}^\nu|, |y_{32}^\nu|, |y_{33}^\nu|\} & =  \{   11.99,  12.57, 
    2.68,  0.50,   0.5, 12.57,  0.5,  1.03,  0.56\}, \nonumber \\
\{\phi_{11}^\nu, \phi_{12}^\nu,  \phi_{13}^\nu,\phi_{21}^\nu,\phi_{22}^\nu,\phi_{23}^\nu, \phi_{31}^\nu,  \phi_{32}^\nu,  \phi_{33}^\nu  \} & =  \{  0.46, 5.22,  2.65,  3,  1.87, 
   6.16,  5.05,  3.90, 
    3.64 \}, \nonumber \\
\epsilon_1  =  8.54 \times 10^{-6},~\epsilon_2  =  3.75 \times 10^{-3},~\epsilon_3  =  0.89,~\epsilon_4 & =  2.28 \times 10^{-5}, 
\epsilon_5  =  4.29 \times 10^{-4},~\epsilon_6  =  1.04 \times 10^{-2}, \nonumber \\
\epsilon&=9.99 \times 10^{-11}, \delta =  1.196, 
\chi^2_{\rm min}  =  4.61,
\end{align}
where $\delta$ is the Dirac $CP$ phase of the CKM matrix.  

\subsection{Scalar potential of the HVM}
\label{scalar_potential}
In the absence of interactions among Higgs doublet and singlet scalar fields, there is a global $SU(7)$  symmetry in the scalar sector such that the   fields  $\chi _i $ of the HVM can be accommodated in its fundamental representation  as shown below, 
\begin{equation}
\chi  = \begin{pmatrix}
\chi _1  \\
\chi _2  \\
\chi _3\\
\chi _4  \\
\chi _5  \\
\chi _6\\
\chi_7
\end{pmatrix}.
\end{equation}  
 Hence, the scalar potential is written as,
\bea 
V &=& \mu^2 \varphi^\dagger \varphi + \lambda (\varphi^\dagger \varphi)^2  +   \mu_1^2 \chi ^\dagger \chi    + \lambda_1  ( \chi^\dagger \chi)^2       
+     \lambda_2  \varphi^\dagger \varphi   \chi^\dagger \chi     + V_1,
\label{SP} 
\eea
where
\bea 
V_1& =&    \rho_1( \chi _1^3 + \chi _1^{\dagger 3}) +   \rho_2( \chi _2^3 + \chi _2^{\dagger 3}) +   \rho_3( \chi _3^3 + \chi _3^{\dagger 3})  +   \rho_4 ( \chi _4^3 + \chi _4^{\dagger 3})  +   \rho_5 ( \chi _5^3 + \chi _5^{\dagger 3})    \\ \nonumber
&+&   \rho_6( \chi _6^3 + \chi _6^{\dagger 3}) +  \rho_7( \chi_7^3 + \chi_7^{\dagger 3}) 
\label{SP1} 
\eea
where the term $V_1$  breaks global  symmetries softly\footnote{ The term $V_1$ may be originated from a gauged symmetry which is broken at a higher scale.  Introducing soft-breaking terms is also justified from the cosmological point since it will avoid the domain-wall problem.}. 

 We parametrize the VEVs of the fields $ \chi_i $  in the following way:
\bea
\langle  \chi _i \rangle = \dfrac{1}{\sqrt{2}} (v_i + s_i + a_i).
\eea
The potential is analysed  in the limit where there is no mixing among the Higgs field and the fields $ \chi _i$\footnote{This is done for purely phenomenological purpose keeping in mind that the discovered Higgs at the LHC is behaving like the SM Higgs boson.    However, an origin of this assumption will be provided when we discuss a TC origin of the HVM.}.  This is equivalent of putting $\lambda_2 =0$.   The parameters $\mu$ and $\rho_i$ can be eliminated  by the following  minimization conditions obtained by imposing $v, v_i$ to be the absolute minimum:
\bea
\label{min}
\mu^2 & = & -\lambda v^2, 
v_1 = -\frac{\sqrt{2} \left( \mu_1^2 +  \lambda_1 \omega^2 \right)}{3 \rho_1}, v_2 = -\frac{\sqrt{2} \left( \mu_1^2 +  \lambda_1 \omega^2 \right)}{3 \rho_2}, v_3 = -\frac{\sqrt{2} \left( \mu_1^2 +  \lambda_1 \omega^2 \right)}{3 \rho_3}, \\ \nonumber 
   v_4 &=&  -\frac{\sqrt{2} \left( \mu_1^2 +  \lambda_1 \omega^2 \right)}{3 \rho_4},
   v_5 = -\frac{\sqrt{2} \left( \mu_1^2 +  \lambda_1 \omega^2 \right)}{3 \rho_5}, v_6 = -\frac{\sqrt{2} \left( \mu_1^2 +  \lambda_1 \omega^2 \right)}{3 \rho_6}, v_7 = -\frac{\sqrt{2} \left( \mu_1^2 +  \lambda_1 \omega^2 \right)}{3 \rho_7},
\eea
where $\omega^2 = v_1^2 + v_2^2 + v_3^2 + v_4^2 + v_5^2 + v_6^2 + v_7^2$. 

Thus we observe a seesaw-like  relation between the VEVs of the fields $ \chi _i$  and the corresponding soft breaking term, i.e.,
\begin{equation}
v_i \propto \frac{1}{\rho_i}.
\end{equation}

The mass matrix of $CP$ even components of the fields $ \chi _i$ is constructed by computing the second derivatives $M_{ij}^2= \dfrac{\partial^2 V}{\partial v_i \partial v_j}$, and is given as,
\begin{equation}
\label{scalars}
\M^2_s=\left(
 \scalemath{0.9}{
\begin{array}{ccccccc}
  \lambda_1 d_{11} -\mu_1^2  & 2 \lambda_1 v_1
   v_4 & 2 \lambda_1 v_1 v_5 & 2 \lambda_1 v_1 v_2 & 2 \lambda_1
   v_1 v_6 & 2 \lambda_1 v_1 v_3 & 2 \lambda_1
   v_1 v_7 \\
2  \lambda_1 v_1 v_4 & -  \lambda_1 d_{22}-\mu_1^2 & 2 \lambda_1 v_4 v_5 & 2 \lambda_1 v_2 v_4 & 2
   \lambda_1 v_4 v_6 & 2 \lambda_1 v_3 v_4  & 2
   \lambda_1 v_4 v_7 \\
2  \lambda_1 v_1 v_5 & 2 \lambda_1 v_4 v_5 & -  \lambda_1  d_{33} -\mu_1^2 & 2 \lambda_1 v_2 v_5 & 2
   \lambda_1 v_5 v_6 & 2 \lambda_1 v_3 v_5  & 2
   \lambda_1 v_5 v_7 \\
 2 \lambda_1 v_1 v_2 & 2 \lambda_1 v_2 v_4 & 2  \lambda_1 v_2
   v_5 & - \lambda_1  d_{44} -\mu_1^2  & 2 \lambda_1
   v_2 v_6 & 2 \lambda_1 v_2 v_3   & 2 \lambda_1
   v_2 v_7 \\
2  \lambda_1 v_1 v_6 & 2 \lambda_1 v_4 v_6 & 2 \lambda_1 v_5
   v_6 & 2 \lambda_1 v_2 v_6 &  -\lambda_1  d_{55} -\mu_1^2 & 2 \lambda_1 v_3 v_6    & 2 \lambda_1 v_6 v_7\\
 2 \lambda_1 v_1 v_3 & 2 \lambda_1 v_3 v_4 & 2 \lambda_1 v_3
   v_5 & 2 \lambda_1 v_2 v_3 & 2 \lambda_1 v_3 v_6 & - \lambda_1  d_{66} -\mu_1^2    & 2 \lambda_1 v_3 v_7  \\
   2  \lambda_1 v_1 v_7 & 2 \lambda_1 v_4 v_7 & 2 \lambda_1 v_5
   v_7 & 2 \lambda_1 v_2 v_7 & 2  \lambda_1 v_6 v_7& 2 \lambda_1 v_3 v_7    & - \lambda_1  d_{77} -\mu_1^2 \\
\end{array}
}
\right).
\end{equation} 
where 
\begin{align}
d_{11} &= (2 v_1^2 -\omega^2), \\ \nonumber
d_{22} &= (\omega^2 - 2v_4^2), \\ \nonumber
d_{33} &= (\omega^2  - 2 v_5^2), \\ \nonumber
d_{44} &= (\omega^2 - 2 v_2^2), \\ \nonumber
d_{55} &= (\omega^2 - 2 v_6^2), \\ \nonumber
d_{66} &= (\omega^2 - 2 v_3^2), \\ \nonumber
d_{77} &= (\omega^2 - 2 v_7^2). 
\end{align}
The  mass square eigenvalues of the mass matrix  are given by, 
\bea
m_{S_{1-6}}^2 &= &  -\mu_1^2 -  \lambda_1 (v_3^2 + v_7^2) + \mathcal{O} (\dfrac{v_{1}^2}{v_3^2}),  \\ \nonumber
m_{S_{7}}^2 &= &  -\mu_1^2 +  \lambda_1 (v_3^2 + v_7^2) + \mathcal{O} (\dfrac{v_{1}^2}{v_3^2}).
\eea
The mass matrix of the $CP$ odd components of the scalar fields is diagonal, and masses of pseudo scalar states are,
\bea
m_{A_{1-7}}^2 & = & \frac{3}{2}(  \mu_1^2 + \lambda_1 \omega^2  )  
\eea
In the phenomenological investigation, the parameter  $\mu_{1}$ can be traded with the mass of either a scalar or a pseudo-scalars. 

We note that the conditions given in the eq. \ref{min} are not sufficient for the absolute minimum.  In addition to this, the Jacobian is needed to be positive which  requires that mass-squared eigenvalues must be positive.  This requirement leads to the condition $\lambda_{1} >0$.

 \subsection{A UV  completion of the HVM}
\label{UVC}
For obtaining a UV completion of the HVM, we employ vector-like fermions which are an active field of research and appears various extensions of the SM\cite{Corcella:2021mdl}-\cite{Abbas:2016xgj,Abbas:2017hzw}.  Thus, we introduce  following  vector-like fermionic fields  transforming under the  SM symmetry  $SU(3)_c \times SU(2)_L \times U(1)_Y$  as,
\begin{eqnarray}
F_{L,R} &: &U_{L,R}^i :   (3,1,\dfrac{4}{3}),
D_{L,R}^{i} :   (3,1,-\dfrac{2}{3}),  
N_{L,R}^i :   (1,1,0), 
E_{L,R}^{i} :   (1,1,-2),
\end{eqnarray}
where $i=1,2,3$.  

The transformations of above fields under the $\mathcal{Z}_2$, $\mathcal{Z}_2^\prime$ and $\mathcal{Z}_2^{\prime \prime}$  symmetries are shown in table \ref{tab2}.
\begin{table}[h]
\begin{center}
\begin{tabular}{|c|c|c|c|}
  \hline
  Fields             &        $\mathcal{Z}_2$                    & $\mathcal{Z}_2^\prime$   & $\mathcal{Z}_2^{\prime \prime}$       \\
  \hline
     $ U_{L,R}^1, N_{L,R}^1, D_{L,R}^1, E_{L,R}^1 $                         & +  &      +   &   -                    \\
     $  U_{L,R}^2, N_{L,R}^2, D_{L,R}^2, E_{L,R}^2 $                         & +  &      -     &  +                   \\
       $  U_{L,R}^3, N_{L,R}^3, D_{L,R}^3, E_{L,R}^3 $                         & -  &      +     &  +                   \\
  \hline
     \end{tabular}
\end{center}
\caption{The charges of vector-like  fermions  under $\mathcal{Z}_2$,  $\mathcal{Z}_2^\prime$  and  $\mathcal{Z}_2^{\prime \prime}$   symmetries.}
 \label{tab2}
\end{table} 

The interactions of vector-like fermions with the SM  fermions are given by,
\begin{eqnarray}
\label{int1}
\mathcal{L} &= & y_u^{i} \bar{\psi}_{Lq}^{i} \tilde{\varphi} U_R^i + y_d^{i} \bar{\psi}_{Lq}^{i} \varphi  D_R^i  + y_\nu^{i} \bar{\psi}_{L\ell}^{i} \tilde{\varphi} N_R^i 
+  y_\ell^{i} \bar{\psi}_{L\ell}^{i} \varphi  E_R^i   +  {\rm H.c.}.
\end{eqnarray}
The  singlet scalar fields interact with vector-like  fermions  and the singlet fermions of the SM through the following Lagrangian,
\begin{eqnarray}
\label{int2}
\mathcal{L} &=&h_i^u \bar{U}_L^i \chi_i \psi_R^u  +   h_i^d \bar{D}_L^i \chi_i \psi_R^d  
+    h_i^\ell \bar{E}_L^i \chi_i \psi_R^\ell       
+    h_i^\nu \bar{N}_L^i \chi_i \psi_R^\nu    
 +  {\rm H.c}, 
\end{eqnarray}
where $u = u_{R}, c_{R}, t_{R}$, $d = d_{R}, s_{R}, b_{R}$, $\ell  = e_{R}, \mu_{R}, \tau_{R}$ and $\nu = \nu_{e_R}, \nu_{\mu_R}, \nu_{\tau_R}$.  

The masses of  new  fermions are derived from the Lagrangian,
\begin{eqnarray}
\label{mass3}
\mathcal{L} &=& M_U^i \bar{U}_L^i  U_R^i + M_D^i  \bar{D}_L^i  D_R^i   
+  M_N^i \bar{N}_L^i  N_R^i  + M_{M}^i \bar{N^c}^{i}_{L,R} N^{i}_{L,R}  +M_E^i  \bar{E}_L^i  E_R^i   + {\rm H.c.}.\qquad
\end{eqnarray}
\section{Technicolour origin of  the  HVM}
\label{TC_model}
In this section, we shall discuss how the hierarchical VEVs can be obtained in a  TC framework.  The well known problem of the standard TC models is the presence of large FCNC effects.  This  is related to the mass generation mechanism of the TC models.  For instance, the mass of an SM fermion is written as,
\be
m_f \propto \frac{\Lambda_{\text{TC}}^3}{\Lambda_{\text{ETC}}^2},
\label{eq2}
\ee
where $\Lambda_{\text{ETC}}$ shows the scale of the extended TC (ETC) gauge sector and $\Lambda_{\text{TC}}$ denotes  the TC mass scale.  For the suppression of the FCNC effects, the $\Lambda_{\text{ETC}}$  must be very high that will result in an unrealistic top quark mass.

The hierarchical VEVs may appear in the framework of dynamical symmetry breaking where the hypothesis of the most attractive channel (MAC) creates favourable energy scales in the form of chiral condensates\cite{Raby:1979my}.  The MAC hypothesis is extended to the multi-fermion case and is referred to as the extended MAC scenario (EMAC)\cite{Aoki:1983ae,Aoki:1983yy}.

In the EMAC scenario, the most attractive channel for the two-body system for an $SU(N)$ gauge theory is the $\bar{\psi}_L \psi_R$ state, and the state $(\bar{\psi}_L \psi_R)^n$ turns out to be the  most attractive channel  as $n$ approaches to a larger value.  The energy  associated with a multi-fermion condensation is given as\cite{Aoki:1983ae,Aoki:1983yy},
\bea
\bar{E} (n) = \dfrac{1}{n} E (\bar{\psi}_L^{n/2} \psi_R^{n/2}) \lesssim V_{E}^{LL} \dfrac{N^2-1}{N} - V_M^{LL} \dfrac{N-1}{N},
\eea
where $V_E^{LL}$ and $V_M^{LL}$ are  the electric and magnetic part of the Hamiltonian of two fermions system.

We observe that the multi fermion systems  become more attractive for the  larger values of $n$ resulting the hierarchical structure for the multi fermion condensations:
\be
\langle \bar{\psi}_L \psi_R \rangle << \langle \bar{\psi}_L  \bar{\psi}_L \psi_R  \psi_R \rangle <<  \langle \bar{\psi}_L \bar{\psi}_L  \bar{\psi}_L \psi_R  \psi_R \psi_R \rangle << \cdots.
\ee
This hierarchy can be parametrized as\cite{Aoki:1983yy}, 
\be 
\label{VEV_h}
\langle  ( \bar{\psi}_L \psi_R )^n \rangle \sim \left(  \Lambda \exp(k \Delta \chi) \right)^{3n},
\ee
where $\Delta \chi$ denotes the chirality of an operator, $k$  a constant and $\Lambda$ shows the scale of the underlying gauge theory.  

This results in the breaking of  an axial $U(1)$ symmetry  in a hierarchical pattern\cite{Aoki:1983yy},  
\be 
U(1)_A \rightarrow \mathcal{Z}_m \rightarrow \cdots \rightarrow  \mathcal{Z}_8  \rightarrow  \mathcal{Z}_6 \rightarrow  \mathcal{Z}_4 \rightarrow  \mathcal{Z}_2, 
\ee
where the first step of the breaking   is caused by the instanton effects.   

Before we begin to construct our TC framework, we recall an important ad hoc assumption which we made while analysing the scalar potential of the HVM. In the section \ref{scalar_potential}, we assumed that the mixing between the Higgs field and the singlet scalar fields must be very small owing to the observed fact that the discovered Higgs boson is behaving like the SM Higgs boson so far.  This is equivalent to saying that $\lambda_2  \approx 0$.  

This fact may be a guiding principle for building models beyond the SM.  Moreover, this may be the first hint that the Higgs field is not a fundamental scalar field.  The reason is that almost all  the beyond the SM theories contain extra scalar fields.  These extra fields always mix with the SM Higgs field.  If these scalar fields are fundamental, then their interaction coupling has to be chosen very small  to make it compatible with the discovered Higgs data without providing a theoretical reason.  However, if these scalar fields are bound states of some new strong forces, then there is a possible mechanism, as we shall show in the following discussion,  to suppress  the mixing between the Higgs field and the additional scalar fields.

We shall now present a TC origin of the HVM by using EMAC scenario with a new mechanism of associating different VEVs to different fermions among and within the families.  This mechanism is inspired by the mechanism discussed in ref.\cite{Harari:1981bs} for   identifying generations of fermions in a different context.
 
As discussed above, the mixing between the SM Higgs field and singlet scalar fields must be very small to satisfy the discovered Higgs data.  This can be achieved if the Higgs field is a bound state of a TC force and singlet scalar fields are bound states of a different TC force.  Moreover, these two different TC forces are accommodated in a common ETC theory.   Then mixing between the Higgs field and singlet scalar fields will be mediated by the ETC gauge sector.  Hence, it is suppressed atleast by the factor $1/ \rm \Lambda_{ETC}^2$ resulting  extremely small values for the quartic coupling  $\lambda_2$.  This is the reason and motivation for us to introduce two TC symmetries $SU(N_{TC})$ and $SU(N_{DTC})$, where DTC stands for ``dark TC"  in our model.  

Moreover, it is assumed that the vector-like fermions $F_{L,R}$ correspond to a different  QCD-like gauge group $SU(N_F)$.   Thus our model is a strongly interacting sector based on $SU(N_{TC)}  \times SU(N_{ DTC}) \times SU(N_{ F})$ symmetry.

Our TC  model contains only one doublet of fermions of Weinberg\cite{Weinberg:1975gm}  and Susskind\cite{Susskind:1978ms} models  transforming under   $SU(3)_c \times SU(2)_L \times U(1)_Y \times SU(N_{TC}) \times SU(N_{DTC}) \times SU(N_F)$ as,
\begin{eqnarray}
T_{q}  &\equiv&   \begin{pmatrix}
T  \\
B
\end{pmatrix}_L:(1,2,0,N_{TC},1,1),  \\ \nonumber
T_{R} &:& (1,1,1,N_{TC},1,1), B_{R} : (3,1,-1,N_{TC},1,1), 
\end{eqnarray}
where electric charges $+\frac{1}{2}$ for $T$ and $-\frac{1}{2}$ for $B$.

For the $\rm DTC$ symmetry our model is,
\begin{eqnarray}
 \mathcal{D}_{ q}^i &\equiv& \mathcal{C}_{L,R}^i  : (1,1, 1,1,N_{DTC},1),~\mathcal{S}_{L,R}^i  : (1,1,-1,1,N_{DTC},1), 
\end{eqnarray}
where $i=1-6$ and  electric charges $+\frac{1}{2}$ for $\mathcal C$ and $-\frac{1}{2}$ for $\mathcal S$.

The symmetry $SU(N_{\text{F}})$ have the following fermions,
\begin{eqnarray}
F_{L,R} &\equiv &U_{L,R}^i \equiv  (3,1,\dfrac{4}{3},1,1,N_F),
D_{L,R}^{i} \equiv   (3,1,-\dfrac{2}{3},1,1,N_F),  \\ \nonumber 
N_{L,R}^i &\equiv&   (1,1,0,1,1,N_F), 
E_{L,R}^{i} \equiv   (1,1,-2,1,1,N_F),
\end{eqnarray}
where $i=1,2,3$.  

It is obvious that we have a global $SU(12)_L \times SU(12)_R \times U(1)$ flavour symmetry corresponding to  the $SU(N)_{DTC}$.  Furthermore, there exits an extra $U(1)_{ X_{DTC}}$ axial symmetry for  the $SU(N_T)_{DTC}$ gauge group.   We further assume that the SM fermions, techniquarks and vector-like fermions can be further embedded in an ETC group $G_{ETC} $  where $ G_{ETC}$  is not known yet.  

Moreover the $U(1)_{X_{DTC}}$ axial  symmetry is broken by instantons of dark TC force  leading  to non vanishing VEV for a $2 K$-fermion operator having $X_{DTC}= 2 K$ quantum number  where $K$ is the number of flavours\cite{Harari:1981bs}. This operator does not have any other quantum number such as colour or flavour resulting  the breaking of the $U(1)_{X_{DTC}}$ axial  symmetry to $\mathcal{Z}_{2K}$ conserved subgroup.  This breaking provides us  conserved axial quantum numbers $X_{DTC}$ modulo $2K$\cite{Harari:1981bs}.  

We can now form composite operators corresponding to the singlet scalar fields $\chi_i$ which can be associated with different SM fermions among and within the generations.  These operators must carry non-vanishing global axial charge $X_{DTC}$ which is conserved modulo $2K$.  These charges of DTC and SM fermions under $U(1)_{X_{DTC}}  $ symmetries are  defined  in table \ref{tab3}.
\begin{table}[h]
\begin{center}
\begin{tabular}{|c|c|}
  \hline
  Fields                &      $U(1)_{X_{DTC}} $            \\
  \hline
       $ \bar{C}_{L}^i, \bar{\mathcal{N}}_{L}^i,  \bar{S}_{L}^i,  \bar{\mathcal{E}}_{L}^i$                            &      1            \\
      $ C_{R}^i, \mathcal{N}_{R}^i, S_{R}^i, \mathcal{E}_{R}^i$                            &     1            \\
         $ \bar{\psi}_{L}^1$                     &    -  2            \\
         $ \bar{\psi}_{L}^2$                    &    -  10            \\
         $ \bar{\psi}_{L}^3$                     &    -  18            \\
          $ \psi_{R}^u$                             &     0          \\
       $  \psi_{R}^d$                            &     4          \\
  \hline
     \end{tabular}
\end{center}
\caption{The additive charges of left- and right-handed SM and DTC fermionic fields under $U(1)_{X_{ DTC}}$  symmetry.}
 \label{tab3}
\end{table} 

Thus the breaking of  the $U(1)_{ DTC}$ axial  symmetry undergoes the following steps,
\be 
U(1)_{\rm X_{DTC}}   \rightarrow  \mathcal{Z}_{24}  \rightarrow  \mathcal{Z}_{22} \rightarrow  \mathcal{Z}_{20} \rightarrow   \mathcal{Z}_{18} \rightarrow  \mathcal{Z}_{16} \rightarrow  \mathcal{Z}_{14}  \rightarrow  \mathcal{Z}_{12} \rightarrow  \mathcal{Z}_{10} \rightarrow  \mathcal{Z}_{8} \rightarrow  \mathcal{Z}_{6} \rightarrow  \mathcal{Z}_{4} \rightarrow  \mathcal{Z}_{2},
\ee
where the first breaking is caused by  the instantons effects.   

For instance, the following hierarchical condensates may be created which are subject to the imposition of  $\mathcal{Z}_2$,  $\mathcal{Z}_2^\prime$  and  $\mathcal{Z}_2^{\prime \prime}$ symmetries, 
\bea
\langle \bar{\mathcal C}_L  \mathcal C_R  \rangle <<   \langle (\bar{ \mathcal C}_L  \mathcal C_R)^3  \rangle <<  \langle(\bar{\mathcal C}_L  \mathcal C_R)^5 \rangle  <<  \langle (\bar{\mathcal C}_L  \mathcal C_R)^7 \rangle  
<<  \langle (\bar{\mathcal C}_L  \mathcal C_R)^9 \rangle.
\eea
We can identify the operator $(\bar{\mathcal C}_L  \mathcal C_R)^9$ having $X_{DTC}=18$ with the  field $\chi_3$, the operator $(\bar{\mathcal C}_L  \mathcal C_R)^7$ having $X_{DTC}=14$ with the  field $\chi_6$,  the operator $(\bar{\mathcal C}_L  \mathcal C_R)^5$  having $X_{DTC}=10 $ with the  field $\chi_2$, the operator $(\bar{\mathcal C}_L \mathcal C_R)^3$  having $X_{DTC}=6 $ with the  field $\chi_5$ and the operator $    \bar{\mathcal C}_L \mathcal C_R    $  having $X_{DTC}=2 $ with the  field $\chi_1$.  The field $\chi_4$ is an operator of the form $\bar{\mathcal C}_R \mathcal C_L $ having $X_{DTC}=-2$.  

Thus,  we have obtained five energy scales, discussed in section \ref{intro},  which can distinguish the SM fermions within and among generations resulting an explanation for the fermionic mass spectrum of the SM.
\begin{figure}[ht]
  \centering
  \includegraphics[width=.6\linewidth]{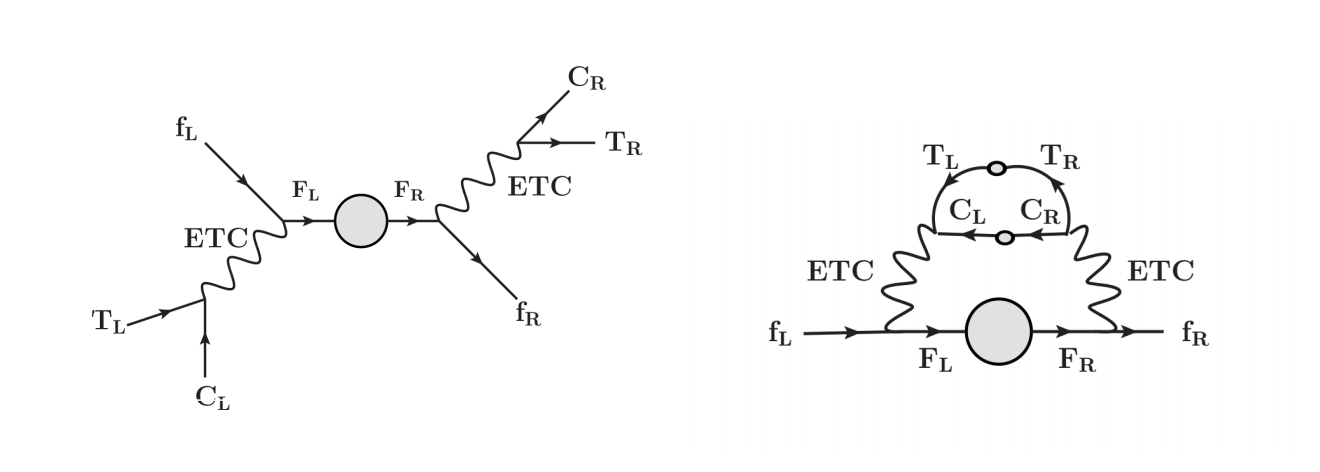}  
  \caption{On the left we show ETC gauge boson interaction with the techiquarks, vector-like fermions and the SM fermions. The contribution to the mass of the $u$-quark  is shown on the right.}
  \label{fig1}
\end{figure}

The creation of dimension-5 operators of the HVM  and mass generation of $u$-quark through them  is shown in fig.\ref{fig1}.   On energy scales much lower than the ETC scale, the following effective six-fermions interaction will be produced,
\be
\bar{\beta}_{ij} \dfrac{\bar{T} \gamma_\mu \bar{d}^i \bar{\mathcal C} \bar{f} f T \gamma^\mu \bar{d}^j \mathcal C}{\Lambda_{\rm ETC}^4 \Lambda_{\rm F}},
\ee
where $\beta$ are coefficients related to the underlying ETC theory.  The chiral factors are hidden in $d^i$s. 

By performing  Fierz rearrangement we obtain,
\be
\beta_{ij} \dfrac{\bar{T} d^i  T \bar{\mathcal C} d^j \mathcal C  \bar{f} f }{\Lambda_{\rm ETC}^4 \Lambda_{\rm F}}.
\ee
The mass of the SM fermion, for instance the mass of the $u$-quark, can be approximated as,
\be
m_f \propto \beta \frac{\Lambda_{\text{TC}}^{3}}{\Lambda_{\text{ETC}}^2}  \dfrac{1}{\Lambda_{\text{F}}} \frac{\Lambda_{\text{DTC}}^{3}}{\Lambda_{\text{ETC}}^2} \exp(2 k),
\label{ETCmass}
\ee
where $\Lambda_{\text{F}}$ is the mass scale of the vector-like fermion.

\begin{figure}[ht]
  \centering
  \includegraphics[width=.6\linewidth]{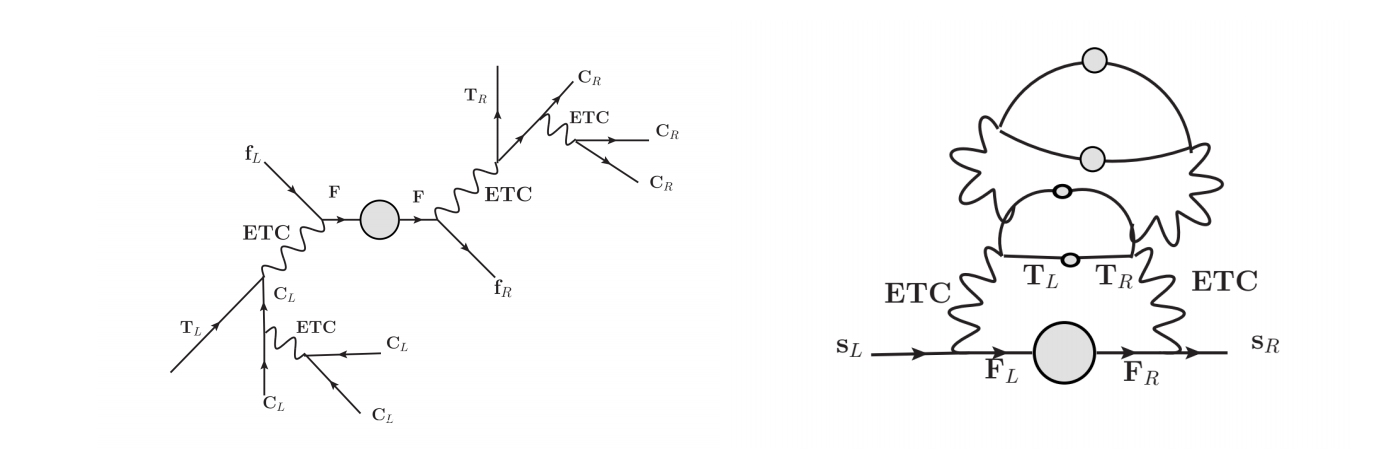}  
  \caption{Interactions of the $s$-quark,  TC and DTC fermions with the ETC gauge sector are shown on the left.  The $s$-quark mass is generated from these interactions  on the right. Here, nameless solid lines represent DTC fermions.}
  \label{fig2}
\end{figure}

In a similar way, one can obtain masses of the other fermions.  In figure \ref{fig2}, we show the interactions which are responsible for providing mass to the $s$-quark.

In summary, the masses of the SM fermions assuming $\beta =1$ can be written using equation \ref{VEV_h}, and are,
\bea
\label{TC_masses}
m_{u} & = & y_u \frac{\Lambda_{\text{TC}}^{3}}{\Lambda_{\text{ETC}}^2}  \dfrac{1}{\Lambda_{\text{F}}} \frac{\Lambda_{\text{DTC}}^{3}}{\Lambda_{\text{ETC}}^2} \exp(2 k),~
m_{c} = y_c \frac{\Lambda_{\text{TC}}^{3}}{\Lambda_{\text{ETC}}^2}  \dfrac{1}{\Lambda_{\text{F}}} \frac{\Lambda_{\text{DTC}}^{11}}{\Lambda_{\text{ETC}}^{10}} \exp(10 k), \\ \nonumber 
m_{t} &=& y_t \frac{\Lambda_{\text{TC}}^{3}}{\Lambda_{\text{ETC}}^2}  \dfrac{1}{\Lambda_{\text{F}}} \frac{\Lambda_{\text{DTC}}^{19}}{\Lambda_{\text{ETC}}^{18}} \exp(18 k), 
m_{d} =  y_d \frac{\Lambda_{\text{TC}}^{3}}{\Lambda_{\text{ETC}}^2}  \dfrac{1}{\Lambda_{\text{F}}} \frac{\Lambda_{\text{DTC}}^{3}}{\Lambda_{\text{ETC}}^2} \exp(2 k), \\ \nonumber
m_{s} &=& y_s \frac{\Lambda_{\text{TC}}^{3}}{\Lambda_{\text{ETC}}^2}  \dfrac{1}{\Lambda_{\text{F}}} \frac{\Lambda_{\text{DTC}}^{7}}{\Lambda_{\text{ETC}}^{6}} \exp(6 k),
m_{b} = y_b \frac{\Lambda_{\text{TC}}^{3}}{\Lambda_{\text{ETC}}^2}  \dfrac{1}{\Lambda_{\text{F}}} \frac{\Lambda_{\text{DTC}}^{15}}{\Lambda_{\text{ETC}}^{14}} \exp(14 k),
\\ \nonumber
m_{e}& = & y_e \frac{\Lambda_{\text{TC}}^{3}}{\Lambda_{\text{ETC}}^2}  \dfrac{1}{\Lambda_{\text{F}}} \frac{\Lambda_{\text{DTC}}^{3}}{\Lambda_{\text{ETC}}^2}\exp(2 k),~
m_{\mu} = y_\mu  \frac{\Lambda_{\text{TC}}^{3}}{\Lambda_{\text{ETC}}^2}  \dfrac{1}{\Lambda_{\text{F}}} \frac{\Lambda_{\text{DTC}}^{7}}{\Lambda_{\text{ETC}}^{6}} \exp(6 k),
m_{\tau} = y_\tau \frac{\Lambda_{\text{TC}}^{3}}{\Lambda_{\text{ETC}}^2}  \dfrac{1}{\Lambda_{\text{F}}} \frac{\Lambda_{\text{DTC}}^{15}}{\Lambda_{\text{ETC}}^{14}} \exp(14 k),
\eea
where we have assumed $\Delta \chi = 0$ for the $TC$ and $F$ sectors.
\subsection{Scalar mass, TC and F scales}
We can have some indication  on the dynamics of the TC and F symmetries  through the  't Hooft large $N$ limit\cite{tHooft:1973alw,Witten:1979kh}.   Using the fact that $SU( \rm N_{\rm F})$ is only a scaled-up QCD, the scale $\Lambda_{\rm F}$   can be related to the QCD scale via following scaling relation,
\bea
\Lambda_{ F} & \sim& \sqrt{\frac{3}{N_{ F}}}  \frac{ F_{ F}}{ F_\pi} \Lambda_{ QCD}.
\eea
Now assuming that $SU(N_{\rm TC})$ is a scaled-down $SU(N)_{\rm F}$, the scale $\Lambda_{\rm TC}$ is given by,
\bea
\Lambda_{ TC} &\sim& \sqrt{\frac{N_{ F}}{N_{ TC}}}  \frac{ F_{ TC}}{ F_{ F}} \Lambda_{ F}.
\eea

Thus, we write the non-trivial scaling relation of our model,
\bea
\label{TCF}
\Lambda_{ F}^2 &\sim& \sqrt{\frac{3 N_{ TC}}{N_{ F}^2}}  \frac{ F_{ F}^2}{\rm F_{ TC} F_\pi } \Lambda_{ TC} \Lambda_{QCD},
\eea
which will be used to determine $\Lambda_{\rm F}$ or $\rm F_{\rm F}$.  This equation can be written in the following form
\bea
\label{TCF2}
\Lambda_{\rm TC}  &\sim& \sqrt{\frac{ N_{\rm F}^2 }{3 N_{\rm TC}}}  \frac{\Lambda_{\rm F}^2}{\rm F_{\rm F}^2}      \frac {F_\pi } { \Lambda_{\rm QCD}} F_{\rm TC}.
\eea
For  $\Lambda_{\rm F} \approx  \rm F_F$,  $\rm N_{TC} = 3$, $\rm N_F = 11$, we observe that the scale $\Lambda_{\rm TC}$ is suppressed by the factor $ \frac {F_\pi } { \Lambda_{\rm QCD}}$.

Such scaling relations are extensively used in literature since the beginning of the TC paradigm\cite{Dimopoulos:1979sp,Dimopoulos:1980yf}.  They  follow from the Nambu-Jona-Lasinio approximation in large $N$ limit\cite{nambu1961}.  


The electroweak VEV and TC decay constant are related by the following scaling rule, 
\be 
\label{TCV}
F_{\rm TC} \sim v \sqrt{\frac{1}{\rm N_D}} \approx 246 \rm GeV,
\ee
where $v = 246$ GeV and $N_D =1$.

The lightest scalar mass (which is the Higgs boson for our TC model) for a QCD-like theory should be of the order of the dynamical mass scale of the theory\cite{nambu1961}.  For instance, in the case of QCD, it is the $\sigma$ meson whose mass is\cite{delbourgo1982},
\be
m_{\sigma} \approx 2 \Lambda_{\rm QCD}.
\ee
For our model, there are two such relations,
\bea
m_{\rm Higgs  } &\approx& 2 \Lambda_{\rm TC}.
\eea
where $ m_{\rm Higgs  } = 125$GeV\cite{atlas,cms}, and 
\bea
m_{\rm \sigma_F  } &\approx& 2 \Lambda_{\rm F}.
\eea
This provides us the following scaling relation,
\bea
m_{\rm Higgs} &\approx&  \frac{ \Lambda_{\rm TC}}{\Lambda_{\rm F}} m_{\rm \sigma_F  }.
\eea
Thus we see that there is $\frac{ \Lambda_{\rm TC}}{\Lambda_{\rm F}}$ suppression in the mass of the Higgs where $ \Lambda_{\rm TC}<<\Lambda_{\rm F}$.

Now the question is what  $\Lambda_{\rm F}$ is.  This can be answered by assuming that the couplings of $ \Lambda_{\rm TC} $ and  $\Lambda_{\rm F}$ are given by a  larger unified theory such as a  GUT with the unification condition,
\be
\alpha_{\rm TC}(\rm M_{GUT}) = \alpha_{\rm F}(\rm M_{GUT}) = \alpha_{3 }( \rm M_{GUT}).
\ee
We can now use the renormalization group evolution to determine the scale $\Lambda_{\rm F}$  to one-loop precision through the following equation\cite{Chetyrkin:2016uhw},
\be 
\frac{\Lambda_{\rm F} }{\Lambda_{\rm QCD} } = \exp\Bigl[   \dfrac{2\pi (\beta_0^{\rm F} - \beta_0^{\rm QCD})}{\beta_0^{\rm F} \beta_0^{\rm QCD} \alpha_{3 }( \rm M_{GUT})}\bigr],
\ee
and
\be 
\frac{\Lambda_{\rm TC} }{\Lambda_{\rm F} } = \exp\Bigl[   \dfrac{2\pi (\beta_0^{\rm TC} - \beta_0^{\rm F})}{\beta_0^{\rm TC} \beta_0^{\rm F} \alpha_{3 }( \rm M_{GUT})}\bigr],
\ee
Thus, the  scale $\Lambda_{\rm F} $ is determined by,
\be 
\frac{\Lambda_{\rm F}^2  }{\Lambda_{\rm TC}  \Lambda_{\rm QCD} } = \exp\Bigl[   \dfrac{2\pi (\beta_0^{\rm F} - \beta_0^{\rm QCD})}{\beta_0^{\rm F} \beta_0^{\rm QCD} \alpha_{3 }( \rm M_{GUT})}\bigr] \exp\Bigl[   \dfrac{-2\pi (\beta_0^{\rm TC} - \beta_0^{\rm F})}{\beta_0^{\rm TC} \beta_0^{\rm F} \alpha_{3 }( \rm M_{GUT})}\bigr]
\ee
where,
\be 
 \beta_0^{\rm QCD} = 11  - \frac{2}{3} n_f,~\beta_0^{\rm TC} = \frac{11 N_T}{3} - \frac{4}{3} N_D, ~ \beta_0^{\rm F} = \frac{11 N_F}{3} - \frac{2}{3} n_{\rm F},
\ee
where $N_D$ are number of doublets and $n_{f,F}$ are number of flavours.

Now using standard choices as inputs we can determine $\Lambda_{\rm F}$.  For instance,  $n_f=6$, $N_T = 3$, $N_D = 1$,  $N_F= 11$, $n_{\rm F} =12 $ and $ \alpha_{3 }( \rm M_{GUT}) = 1/30$, we obtain
\be 
\frac{\Lambda_{\rm F}^2  }{\Lambda_{\rm TC}  \Lambda_{\rm QCD} }  = 3.57417 \times 10^{14}
\ee
Thus for $ \Lambda_{\rm TC} = m_{\rm Higgs  }/2 $ and $\Lambda_{\rm QCD} = 332$ MeV\cite{Tanabashi:2018oca},   we  find $\Lambda_{\rm F} = 8.68048 \times 10^7$ GeV.  

For  $ F_\pi = 130$MeV\cite{Tanabashi:2018oca},   $v=246$ GeV,   and $F_{\rm TC} = v$,  we estimate $F_F = 2.06352 \times 10^8$ GeV from equation \ref{TCF}.

At this point, we naively estimate the asymptotic behaviour of our model.  For this purpose, we assume a standard QCD-like fermionic self-energy  which is\cite{Lane:1974he},
\be 
\frac{1}{4} \rm tr\left(  \Sigma^f (p) \right) = \frac{\Lambda^3}{p^2} \left(  \frac{p}{\Lambda} \right)^{\gamma_m},
\ee
where the anomalous mass dimension is,
\be 
\gamma_m (\mu) = \frac{2 C_2 (R) }{2 \pi} \alpha_s (\mu) + \mathcal{O}( \alpha_s (\mu)^2),
\ee
and $C_2 (R)  $  is the quadratic Casimir of the underlying non-Abelian gauge symmetry, and is given by $C_2 (R)  = \frac{N^2-1}{2N}$ for an $SU(N)$ group.  From this we can estimate the asymptotic behaviour of the masses of the SM fermions which are given in equation \ref{TC_masses}.  For TC sector, we have only one doublet, i.e., only two flavours.  Therefore the anomalous mass dimension $\gamma_m (\mu)<< 1$ when $\Lambda \geq \Lambda_{\rm TC}$ providing,
\be 
\frac{1}{4} \rm tr\left(  \Sigma^f (p) \right)_{TC} \approx  \frac{\Lambda_{\rm TC}^3}{p^2}.
\ee
For QCD we have now 12 flavours (six ordinary quarks and six vector-like quarks from $SU( \rm N_{\rm F})$ sector).  However, the anomalous mass dimension is still  $\gamma_m (\mu)<< 1$ for $\Lambda \geq \Lambda_{\rm QCD}$.  Thus providing negligible contribution to the masses of ordinary fermions.  The same argument is also applicable for DTC sector which contains only 12 flavours.   The scale  $SU( \rm N_{\rm F})$ and $ETC $, as we shall show, are approximately same.  Hence, the contribution to the fermionic self-energy from  the  $SU( \rm N_{\rm F})$ sector does not affect the masses of ordinary fermions at $\Lambda=\Lambda_{\rm F}$ since  $\gamma_m (\mu)<< 1$ for $\Lambda \geq \Lambda_{\rm F}$.  Thus we see that the anomalous mass dimension is always $<< 1$ in our model escaping a walking type behaviour and showing a QCD-like dynamics.\footnote{A more sophisticated calculation requires solving Schwinger-Dyson equation.  This is beyond the scope of this work, and will be presented in a future work. }
\subsection{Neutrino masses }
\begin{figure}[ht]
  \centering
  \includegraphics[width=.6\linewidth]{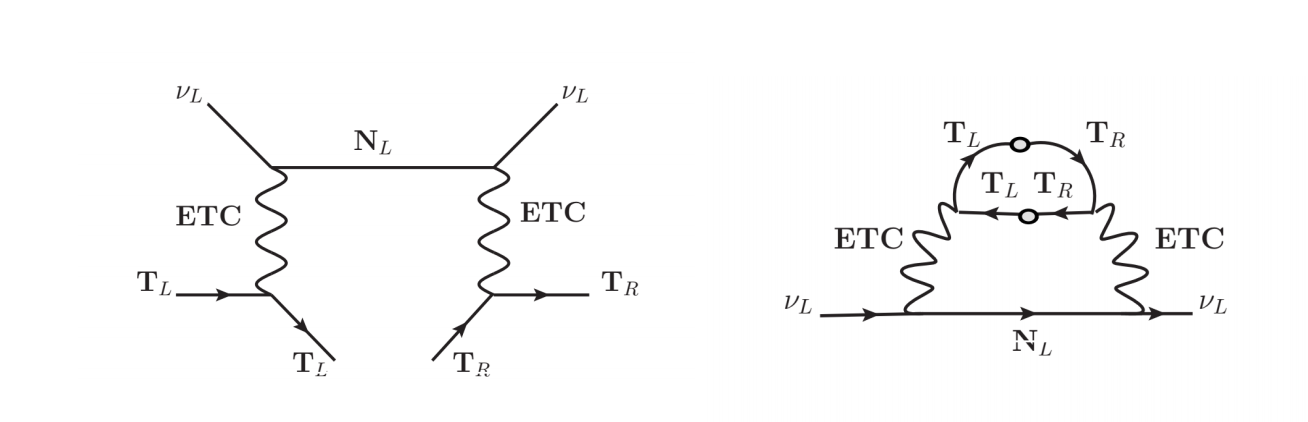}  
  \caption{On the left we show ETC gauge boson interaction with the $TC$ fermions $T_{L,R}$ and the left-handed neutrinos. The contribution to the Majorana mass of the left-handed neutrinos  is shown on the right.}
  \label{left_majo}
\end{figure}

Now we discuss the Majorana mass scales.  In section \ref{neutrino}, we ignored the Majorana mass terms for the left-handed neutrinos assuming it to be much smaller in comparison to that for the right-handed neutrinos.  This can be justified in the present TC model.  For instance, the origin of this term is shown in figure \ref{left_majo}.

\begin{figure}[ht]
  \centering
  \includegraphics[width=.6\linewidth]{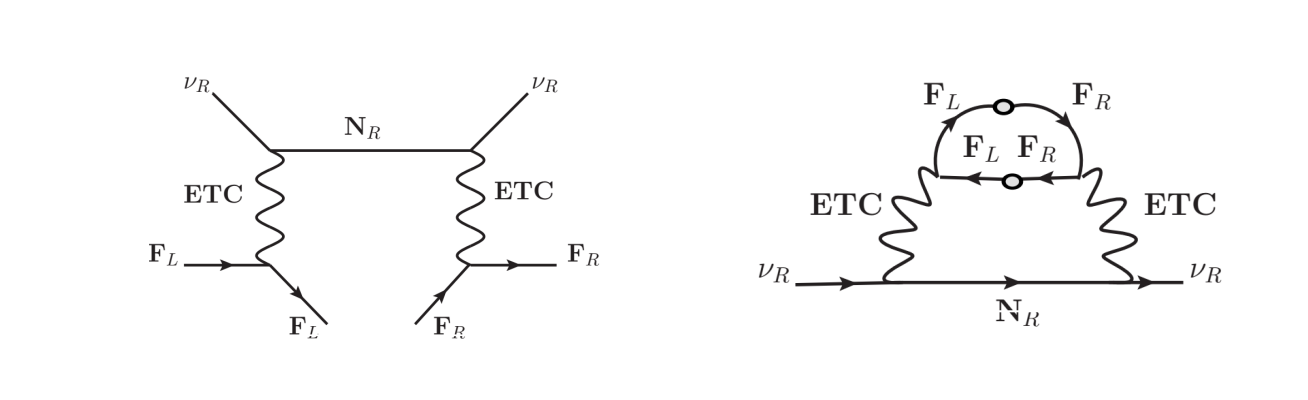}  
  \caption{On the left we show ETC gauge boson interaction with the vector-like fermions $F_{L,R}$ and the right-handed neutrinos. The contribution to the Majorana mass of the right-handed neutrinos  is shown on the right.}
  \label{majorana}
\end{figure}

The contribution to the Majorana mass of the left-handed neutrinos can be approximately written as,
\be 
M_L \propto \frac{\Lambda_{TC}^{6}}{\Lambda_{F} \Lambda^{4}_{ETC}}.
\ee
The diagram which creates the Majorana mass for the right-handed neutrinos  is shown in figure \ref{majorana}. 

The majorana mass is produced by the state  $ (\bar{F}_L  F_R)^{2}  $ which can be identified with the field $\chi_7$.   Thus, the Majorana mass is,
\be 
M_R \propto \frac{\Lambda_{F}^{6}}{\Lambda_{F} \Lambda^{4}_{ETC}},
\ee
where  $\langle \chi_7 \rangle =\langle (\bar{F}_L  F_R)^{2} \rangle $.

The masses of light neutrinos can be written as,
\bea 
\label{TC_neutrino}
m^{\nu}_1& = & y^{\nu}_1 \frac{\Lambda_{\text{TC}}^{3}}{\Lambda_{\text{ETC}}^2}  \dfrac{1}{\Lambda_{\text{F}}} \frac{\Lambda_{\text{DTC}}^{3}}{\Lambda_{\text{ETC}}^2} \exp(2 k) \frac{1}{M_R},~
m^{\nu}_2 = y^{\nu}_2 \frac{\Lambda_{\text{TC}}^{3}}{\Lambda_{\text{ETC}}^2}  \dfrac{1}{\Lambda_{\text{F}}} \frac{\Lambda_{\text{DTC}}^{11}}{\Lambda_{\text{ETC}}^{10}} \exp(10 k)  \frac{1}{M_R}, \\ \nonumber 
m^{\nu}_3 &=& y^{\nu}_3 \frac{\Lambda_{\text{TC}}^{3}}{\Lambda_{\text{ETC}}^2}  \dfrac{1}{\Lambda_{\text{F}}} \frac{\Lambda_{\text{DTC}}^{19}}{\Lambda_{\text{ETC}}^{18}} \exp(18 k)  \frac{1}{M_R}.
\eea
\subsection{ETC scale, Fermion masses, DTC scale and Majorana mass scales}
Flavour changing neutral current (FCNC) processes can be used to estimate the ETC scale $\Lambda_{\rm ETC}$.  For instance, $\Delta S = 2 $ FCNC processes place severe bounds on the ETC scale $\Lambda_{\rm ETC}$ through the $K_L-K_S$ mass difference.  After imposing this constraint, the  ETC scale $\Lambda_{\rm ETC}$ turns out to be\cite{Hill:2002ap},
\be 
\Lambda_{\rm ETC} \gtrsim 10^3 \rm{TeV}.
\ee
We can use the fact that our model should reproduce the fermions masses given in equation \ref{xing2007} to determine the scale $\Lambda_{\rm ETC}$  and $\Lambda_{\rm DTC}$  with the help of equation \ref{TC_masses}.  This is done by assuming  $\Lambda_{\rm F} = 8.68048 \times 10^7$ GeV and $|y_{u,d,c,s,t,b,e,\mu,\tau}| \in [0.5,4 \pi]$.  The results of the fit are,
\bea 
y_u &=& 0.59, y_c = 2.19, 
   y_t = 4.85, y_d =  1.33, y_s =  2.21, 
   y_b =  0.89, y_e =  0.5, \\ \nonumber 
   y_\mu &=&  4.93, 
   y_\tau =  0.65, y_1^\nu =  12.27, y_2^\nu = 12.11, 
   y_3^\nu =  0.54, k  = -13.9886, \\ \nonumber
   \Lambda_{\rm ETC} &=& 1.1 \times 10^{7} \rm GeV, ~ \Lambda_{\rm DTC} = 2.4 \times 10^{13} \rm GeV,    \\ \nonumber 
    M_R &= & 3.36626 \times 10^{11} \rm GeV,
   \chi^2_{\rm min} = 0.053. 
\eea
From above results, we obtain $M_L \approx 4.68992 \times 10^{-17} \rm eV$.  Thus we see that the  Majorana mass scale for the left-handed neutrinos is extremely small, and  can be ignored as we assumed in section \ref{neutrino}.

 \subsection{$TC$ and $DTC$ mixing}
\begin{figure}[htb]
  \centering
  \includegraphics[width=.6\linewidth]{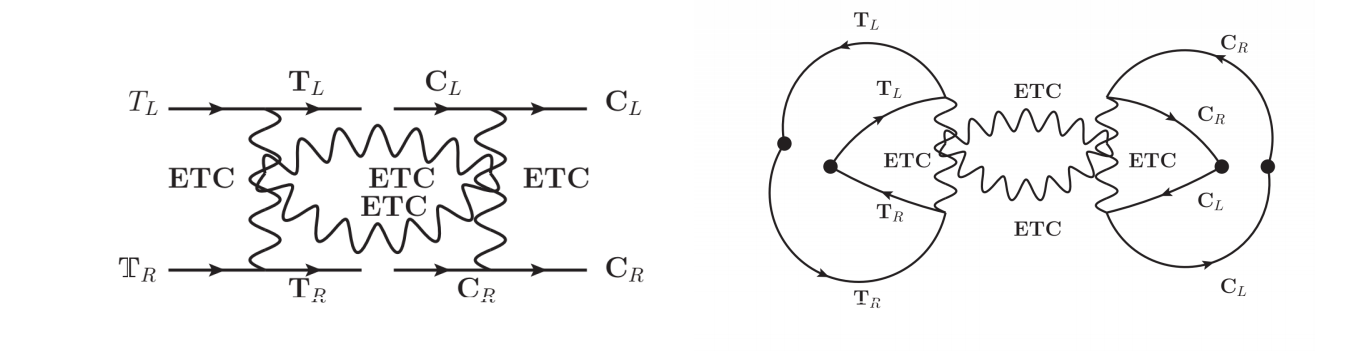}  
  \caption{On the left  eight-fermions interactions mediated by ETC neutral and charged gauge bosons are shown.  We show the creation of the term $\varphi^\dagger \varphi   \chi _1^\dagger \chi _1$ on the right.}
  \label{fig3}
\end{figure}

Now we shall show that the interaction between the Higgs doublet and the singlet scalar fields should be suppressed in our model.  For instance, the term $\varphi^{  \dagger }\varphi   \chi _1^\dagger \chi _1$ can be created through the diagram shown in fig.\ref{fig3}.

The quartic coupling $ \lambda_{\varphi^\dagger \varphi \chi_1^\dagger \chi_1} $ approximately is,
\be
\lambda_{\varphi^\dagger \varphi \chi_1^\dagger \chi_1}\approx \frac{\Lambda_{{\rm TC}}^6  \Lambda_{{\rm DTC}}^6}{\Lambda_{\rm ETC}^{12}} \exp(4 k).
\ee
Using the already determined values of the scales $\Lambda_{\rm TC}$, $\Lambda_{\rm ETC}$, $k$ and $\Lambda_{\rm DTC}$, we find $\lambda_{\varphi^\dagger \varphi \chi_1^\dagger \chi_1} = 1.81614 \times 10^{-18}$.  In a similar manner, we can show that the mixing of the SM Higgs field with other $DTC$ multi fermion condensates are highly suppressed.
 \subsection{$TC$ and $F$  mixing}
We investigate now the interactions between the $TC$ and $F$ sectors.  This can be done through  the term $\varphi^{  \dagger }\varphi   \chi _7^\dagger \chi _7$.  We write this mixing approximately in the following form,
\be
\lambda_{\varphi^\dagger \varphi \chi_7^\dagger \chi_7} \approx \frac{\Lambda_{{\rm TC}}^6  \Lambda_{{\rm F}}^6}{\Lambda_{\rm ETC}^{12}} .
\ee
Using values of the scales $\Lambda_{\rm TC}$, $\Lambda_{\rm ETC}$, and $\Lambda_{\rm F}$, we find $\lambda_{\varphi^\dagger \varphi \chi_7^\dagger \chi_7}  = 8.12513 \times 10^{-27}$.
\subsection{Experimental constraints}
\subsubsection{$S$ parameter}
The obliques parameters place bounds on TC theories\cite{Peskin:1990zt,Peskin:1991sw}. In particular, the  $S$ parameter in TC models may be in conflict with experimental observation.  The experimental values of $S$ and $T$ parameters are \cite{pdg20},
\be  
S=0.00\pm 0.07,~~ T= 0.05\pm 0.06.
\ee
For strongly coupled  theories, these parameters at  next-to-leading-order  are given by\cite{Pich:2013fea},
\begin{align}
S &=4\pi F_\Pi^2 \left[ \frac{1}{M_V^2}+\frac{1}{M_A^2} \right] + \frac{1}{12 \pi} \left[\left( 1-\frac{M_V^4}{M_A^4}\right) \left(   \log \frac{M_V^2}{M_{\rm Higgs}^2} -\frac{11}{6}  \right) + \left( \frac{M_V^2}{M_A^2}-\frac{M_V^4}{M_A^4}\right)  \log \frac{M_A^2}{M_V^2}  \right], \\ \nonumber
T & = \frac{3}{16 \pi \cos^2\theta_W} \left[ (1-\kappa_W^2)      \left(  1+ \log\frac{M_{\rm Higgs}^2}{M_V^2}  - \kappa_W^2   \log\frac{M_V^2}{M_A^2}     \right)                \right],
\label{s1}
\end{align}
where the $\kappa_W = \frac{M_V^2}{M_A^2}$ parametrizes the coupling of the lightest scalar (Higgs boson ) to two gauge bosons ($W^+W^- ~\rm or  ~ZZ$).  These results are generic in nature and, can be used in specific strongly coupled theories such as presented in this work.

In the conventional TC models, the TC dynamics is a scaled up version of the QCD dynamics.  Therefore, the masses of the vector and axial vector bosons $M_{V,A}$ only depend on the TC  and the QCD dynamics (mass of the QCD $\rho$ meson).  Hence, they cannot be very heavy, and  provide a large value of the  $S$ parameter which is difficult to reconcile with experimental observation \cite{Lane:1993wz}.  In the model presented in this work, the TC sector is not a scaled up QCD.  Instead, it is a scaled down version of the $SU(N)_{\rm F}$.  This means that the masses of the vector and axial vector bosons $M_{V,A}$ depend on the $SU(N)_{\rm F}$ dynamics (mass of the $SU(N)_{\rm F}$  $\rho$ meson), and can be much heavier than the conventional TC models, thus, providing a small value of the $S$ parameter which can be in agreement with the experimental value.

Now our objective is to determine the vector meson mass in our TC model.  This can be done by observing the following scaling relation  for the $F$ vector meson\cite{Dimopoulos:1979sp,Dimopoulos:1980yf},
\be 
m_{\rho_{\rm F}} \approx \sqrt{\frac{ 3 }{N_{\rm F}}}  \frac{\rm F_F }{F_\pi} m_{\rho_{QCD}}.
\ee
In a similar manner, we can write,
\be 
m_{\rho_{\rm TC}} \approx \sqrt{\frac{ N_{\rm F} }{N_{\rm TC}}}  \frac{\rm F_{TC} }{F_{F}} m_{\rho_{\rm F}}.
\ee
Now we can write,
\be 
M_V = m_{\rho_{\rm TC}} \approx \sqrt{\frac{ N_{\rm F}^2 }{3 N_{\rm TC}}}  \frac{\rm F_\pi }{m_{\rho_{QCD}}} \frac{m_{\rho_F}^2}{\rm F_F^2} F_{\rm TC},
\ee
where $ F_\Pi = F_{TC} $.

\begin{figure}[ht]
  \centering
  \includegraphics[width=.6\linewidth]{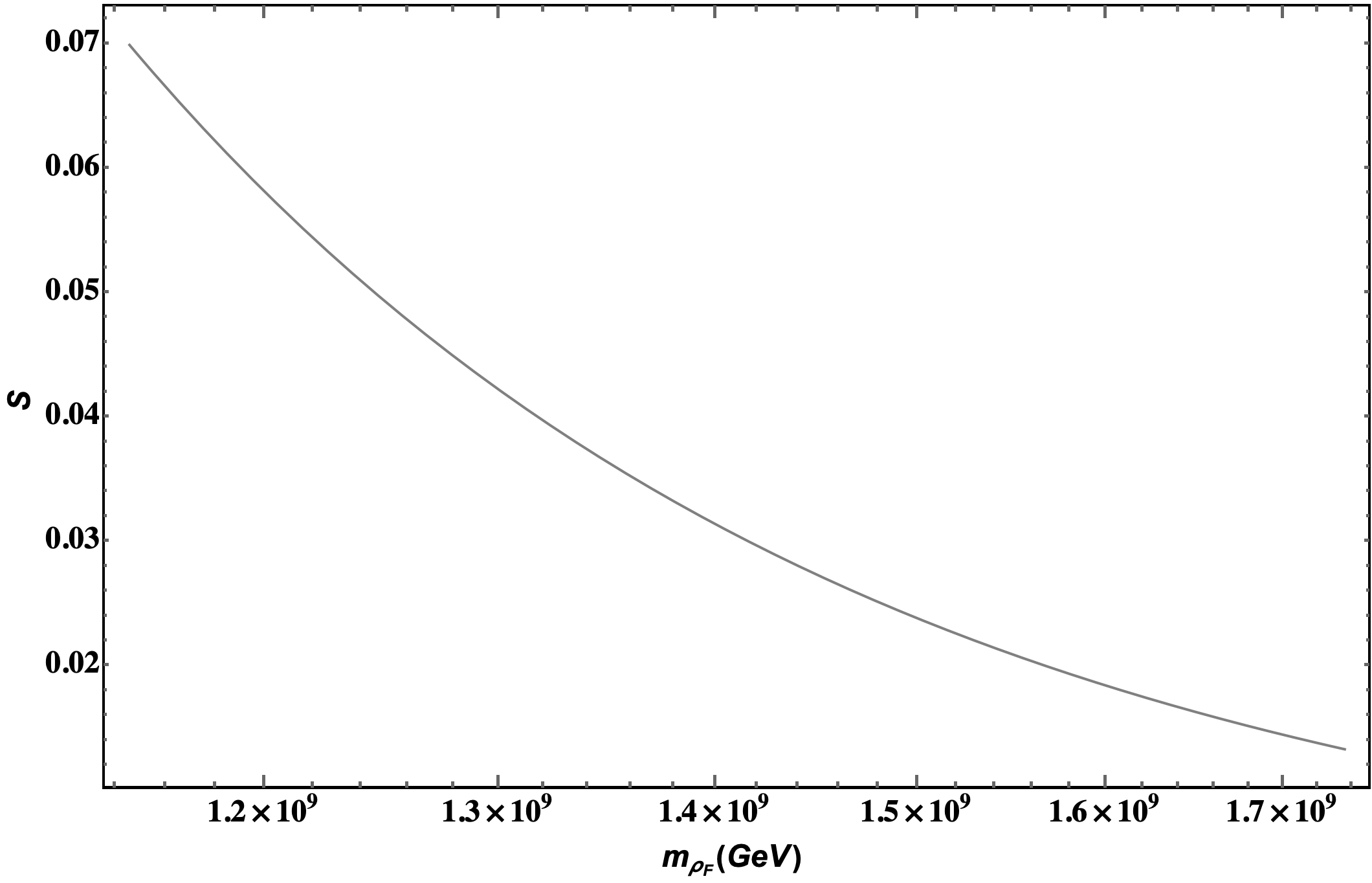}  
  \caption{The variation of the $S$ parameter as a function of the mass of the vector meson $m_{\rho_{F}}$ for a bench mark value $\kappa_W=1$ as adopted in reference \cite{Pich:2013fea}.}
  \label{s_rhf}
\end{figure}

In figure \ref{s_rhf}, we show the allowed range of the $S$ parameter for $m_{\rho_{QCD} }= 775$MeV and a mass range of the vector meson above $m_{\rho_{F}} = 13.2 \Lambda_F - 20  \Lambda_F$ using a bench mark value $\kappa_W=1$ as adopted in reference \cite{Pich:2013fea}.  For this mass range of  $m_{\rho_{F}}$,  we obtain a  range of  the $S$ parameter to be $0.01326-0.07 $.

\begin{figure}[ht]
  \centering
  \includegraphics[width=.6\linewidth]{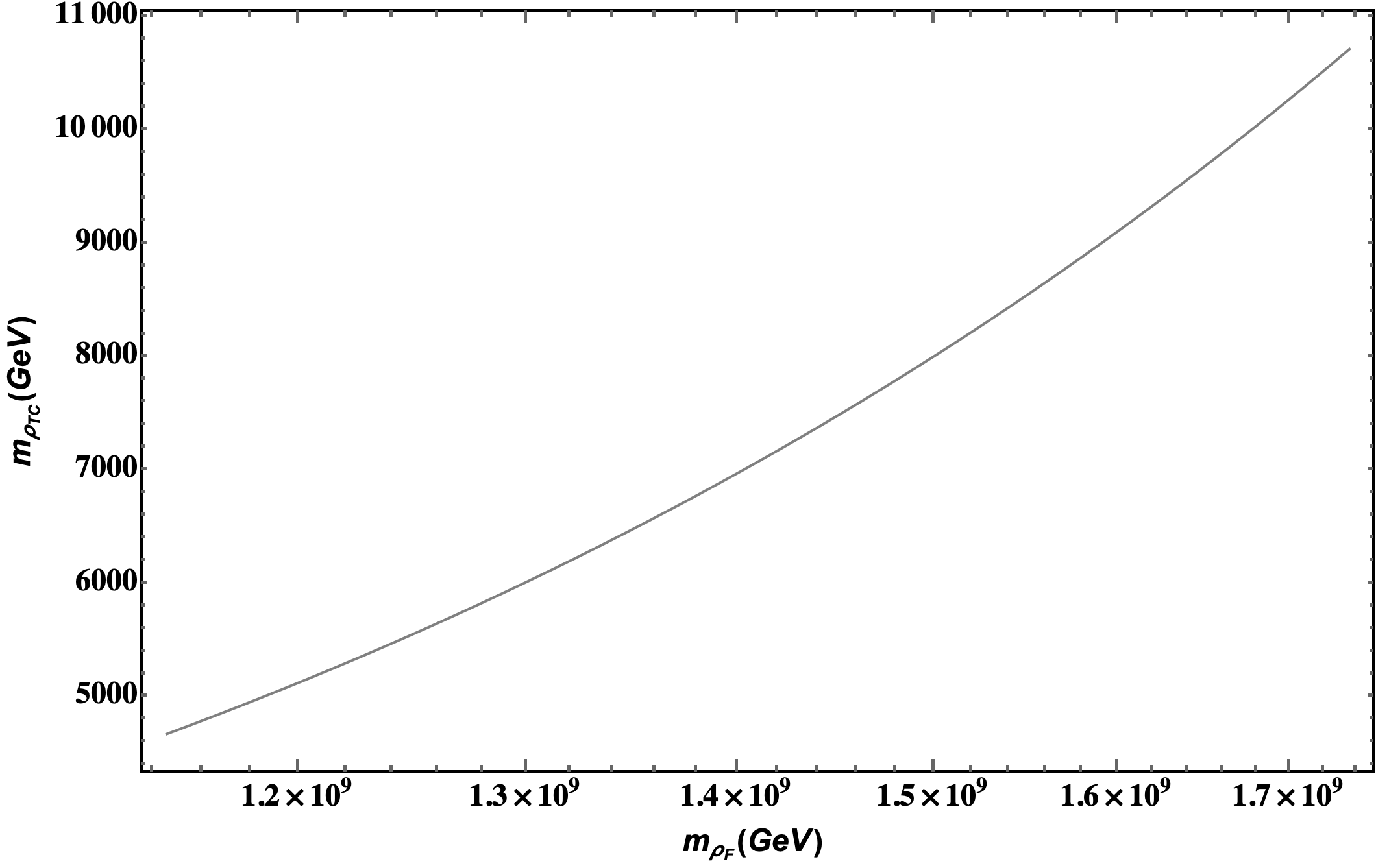}  
  \caption{The variation of the mass of the vector meson  $m_{\rho_{TC}}$ as a function of the mass of the vector meson $m_{\rho_{F}}$.}
  \label{rht_rhf}
\end{figure}

In a similar manner, we determine the variation in the mass of the $TC$ vector meson  $m_{\rho_{TC}}$ as a function of the mass of the vector meson $m_{\rho_{F}} $ which is shown in figure \ref{rht_rhf}.  The mass range of the vector meson  $m_{\rho_{TC}}$ is approximately $4.665- 10.71$TeV for the $m_{\rho_{F}} = 12.3 \Lambda_F - 20  \Lambda_F$.
There are direct and indirect searches  available for this type of vector bosons, and they exclude it mass below $\approx 1$TeV at the Large Hadron Collider \cite{Pich:2013fea,Matsedonskyi:2015dns,CMS:2018ubm,ATLAS:2019nat}.  
\subsubsection{Pseudo-Nambu-Goldstone bosons masses}
The masses of colourless  pseudo-Nambu-Goldstone  bosons  (PNGB) with electroweak gauge charges originate from the gauge interactions, and can be written as,
\be 
m_{\Pi_{ TC}}^2 \sim \frac{F_{ TC}^2 \delta m_\pi^2}{F_\pi^2 \sin^2 \theta_w},
\ee
where $ \delta m_\pi =4.59359 $ MeV.  This  can provide $m_{\Pi_{\rm TC}} \sim 100$GeV. Moreover, we can also write the following scaling relation for the TC PNGB meson,
\be 
m_{\Pi_{\rm TC}} \approx \sqrt{\frac{ N_{\rm F}^2 }{3 N_{\rm TC}}}  \frac{\rm F_\pi }{m_{\pi_{QCD}}} \frac{m_{\Pi_F}^2}{\rm F_F^2} F_{\rm TC}.
\ee
which can also give $m_{\Pi_{\rm TC}} \sim 100$GeV by assuming $m_{\Pi_F} \approx 0.83 \Lambda_F$.
\subsubsection{Scalar boson trilinear coupling}
The measurement of the discovered Higgs trilinear self coupling is an important test to determine whether it is a fundamental or a composite boson\cite{Doff:2021npl}.  
The SM trilinear Higgs boson self-coupling can be extracted from the Lagrangian~\cite{chang_luty},
\be
{\cal{L}}^{\rm SM}_{ H^3}=\lambda^{\rm SM}_{ H^3}  H^3 \, 
\ee
where 
\be
\lambda^{\rm SM}_{ H^3}=\frac{ m_H^2}{2v} =0.129 \, .
\ee
The trilinear self-coupling of the observed Higgs boson can be parametrized through the following Lagrangian\cite{Doff:2021npl},
\be
{\cal{L}}_{ H^3}= \xi  \lambda^{\rm SM}_{ H^3} v H^3 \, ,
\ee
where 
\be
\xi = \frac{\lambda_{ H^3}}{\lambda_{\ H^3}^{\rm SM}} \,\, ,
\ee
where $\xi$ quantifies the modification of the trilinear Higgs boson self-coupling.  The bound on $\xi$  is given by the CMS Collaboration  at $95\%$ CL~\cite{cms_2b2g}
\be
-3.3 < \xi < 8.5.
\ee

An estimate of the trilinear self-coupling of the Higgs boson can be obtained from\cite{Doff:2021npl},
\be 
\lambda_{ H^3} = \frac{1}{16\pi^2}\left(\frac{3N_{F}}{F^4_{TC}}\right)\int^{\Lambda^2_{ETC}}_{0}\frac{\Sigma^4(p^2)p^4dp^2}{(p^2 + \Sigma^2(p^2))^3},
\label{tri3} 
\ee
where $N_F$ are number of $TC$ fermions and, 
\be
\Sigma (p^2) \approx \frac{\Lambda_{TC}^3}{p^2}.
\label{eq1}
\ee
Thus, for our numerical inputs, we obtain,
\be
\lambda_{ H^3} = 2.127 \times 10^{-5},
\ee
which translates into 
\be 
\xi = 1.164883  \times 10^{-4}.
\ee

\section{ Summary and discussion}
\label{sum}
The flavour problem of the SM is one of the most challenging puzzles which has potential to lead to the physics beyond the  SM.  In this work, we address this problem through an unconventional approach based on the VEVs hierarchy \cite{Abbas:2017vws}.  An origin of the VEVs hierarchy is presented in a conventional non-minimal TC    model where the SM Higgs boson is a bound state of the conventional TC gauge theory, and the gauge singlet scalar fields $\chi_i$ originate from a dark TC dynamics.   We note that the fermions masses are generated mainly through  the DTC sector.  

The Majorana mass of the right-handed neutrinos  has origin in a different strong sector denoted by $F$, and due to this the model is capable of explaining neutrino masses and mixing angles.  The mixing between $TC$ and $F, DTC$ sectors is highly suppressed, providing an explanation for the observed behaviour of the discovered Higgs boson.  Moreover, the essential experimental constraints are also satisfied by this non-minimal TC framework.

For the  multi-fermions condensates which play the role of the gauge singlet scalar fields $\chi_i$, it is assumed that the three discrete symmetries $\mathcal{Z}_2$,  $\mathcal{Z}_2^\prime$  and  $\mathcal{Z}_2^{\prime \prime}$ have an independent origin, for instance, in three vectorial $U(1)_{\rm V}$ symmetries present in the model. 

We have not discussed an explicit ETC model for the non-minimal TC  model of this work  due to its non-trivial nature which requires an in-depth investigation.  However, we observe that in the non-minimal TC  model of this work, construction of an ETC model may be simpler than the conventional ETC models.  In the conventional ETC models, the symmetry of the model is expected to be broken in several steps so that an explanation for the hierarchical fermion masses among generations  may be achieved.  In the  non-minimal TC  model presented in this work, an origin as well as an explanation for the hierarchical fermion masses arise from a DTC sector.  Therefore, there is no need to break the ETC symmetry in several steps.

\section*{Acknowledgements}
I am extremely grateful to Prof. Anjan S. Joshipura for very helpful discussion on the HVM in the past.  Feynman diagrams are created through the package JaxoDraw\cite{Binosi:2003yf}.

\begin{appendix}
\section{Outline of a possible ETC model}
We comment here now about an important implication of the model.  The vector-like fermions are assumed to be  associated with the group  $SU( \rm N_{\rm F})$ whose scale is $\Lambda_{\rm F}$.  We have chosen this group for the present work.  However, it could be any other symmetry.  It turns out that the ETC scale and the scale $\Lambda_{\rm F}$ are approximately same in our model.  Therefore, the symmetry $SU( \rm N_{\rm F})$ can be identified with that of the ETC symmetry, and vector-like fermions $F_{L,R}$ are in fact ETC fermions.  This observation can help in guessing a possible outline of an ETC model.  

Considering above observation, the symmetry of our model becomes a minimal symmetry, i.e.   $SU(3)_c \times SU(2)_L \times U(1)_Y \times SU(N_{TC}) \times SU(N_{DTC})$  where TC sector is,
\begin{eqnarray}
T_{q}  &\equiv&   \begin{pmatrix}
T  \\
B
\end{pmatrix}_L:(1,2,0,N_{TC},1),  \\ \nonumber
T_{R} &:& (1,1,1,N_{TC},1), B_{R} : (3,1,-1,N_{TC},1), 
\end{eqnarray}
where electric charges $+\frac{1}{2}$ for $T$ and $-\frac{1}{2}$ for $B$.

We note that the ETC gauge boson should interact with TC and DTC sector simultaneously so that maases of the SM fermions can be recovered in the model.  For this reason, the $\rm DTC$ fermions are also charged under TC sector,
\begin{eqnarray}
 \mathcal{D}_{ q}^i &\equiv& \mathcal{C}_{L,R}^i  : (1,1, 1,N_{TC},N_{DTC}),~\mathcal{S}_{L,R}^i  : (1,1,-1,N_{TC},N_{DTC}), 
\end{eqnarray}
where $i=1-6$ and  electric charges $+\frac{1}{2}$ for $\mathcal C$ and $-\frac{1}{2}$ for $\mathcal S$.  The asymptotic freedom in the TC sector can be ensured by taking a large value of $N_{TC}$ such that the anomalous mass dimension $\gamma_m (\mu)<< 1$ when $\Lambda \geq \Lambda_{\rm TC}$.

The vector-like fermions $F_{L,R}$ are now belong to ETC sector which is represented by the group $SU(N_{\text{F}})$ in the model.   These are,
\begin{eqnarray}
F_{L,R} &\equiv &U_{L,R}^i \equiv  (3,1,\dfrac{4}{3},1,1,N_F),
D_{L,R}^{i} \equiv   (3,1,-\dfrac{2}{3},1,1,N_F),  \\ \nonumber 
N_{L,R}^i &\equiv&   (1,1,0,1,1,N_F), 
E_{L,R}^{i} \equiv   (1,1,-2,1,1,N_F),
\end{eqnarray}
where $i=1,2,3$.  

Thus,  the underlying symmetry of our model  may be of the type,
\be 
  SU(\rm N_{\rm DTC}) \times  G_{\rm ETC},
\ee
where $G_{\rm ETC}$ contains the SM and TC symmetries in such a way that the SM, TC fermions (which include DTC fermions as well)  live in a single irreducible representation of $G_{\rm ETC}$.  For instance,
\begin{eqnarray}
 \psi_{\rm ETC}  &\equiv&\rm    \begin{pmatrix}
f  \\
T (\mathcal{D})
\end{pmatrix},
\end{eqnarray}
where $f$ represents the SM fermions.  $G_{\rm ETC}$ symmetry breaks down to the  TC sector   at the energy scale $\Lambda_{\rm ETC}$. A detailed investigation of the $G_{\rm ETC}$ symmetry is beyond the scope of this work, and  will be studied in a future work.

\section{An alternative HVM }
We discuss a scenario where we have only three  fields $F_i$ to create three different energy scales to account for the mass hierarchy among the fermionic families. The mass hierarchy within the  fermionic families can be recovered from the electroweak scale itself if we perform electroweak symmetry breaking through the two Higgs doublets where the first doublet only couples to the $down$ type fermions and the  second doublet couples to the $up$ type fermions.  Thus we have  a type \rom{2} two Higgs doublet model extended by four gauge singlet scalar fields.

We begin with  four singlet scalar fields $\chi _{i}$  and $\chi _1^\prime$ transforming   under the SM symmetry  $SU(3)_c \times SU(2)_L \times U(1)_Y$ as,
\begin{eqnarray}
\chi _i, \chi_1^\prime :(1,1,0),
 \end{eqnarray} 
where $i=1,2,3$ in general, and two SM Higgs doublets $\varphi_1$ and $\varphi_2$.   

In addition to this,  we add three discrete $\mathcal{Z}_2$   symmetries to the SM as described  in the table  \ref{tab_alt}.  The  masses of fermions now originate from dimension-5 operators which are,
\bea
\label{massAp}
\mathcal{L} &=& \dfrac{1}{\Lambda}   \Bigl(    y_{ij}^u  \bar{\psi}_L^{q,i}  \tilde{\varphi_2} \psi_R^{u, j}   \chi _i +  y_{ij}^d  \bar{\psi}_L^{q,i}   \varphi_1 \psi_R^{d, j}  \chi _{i}   +   y_{ij}^\ell  \bar{\psi}_L^{\ell, i}   \varphi_1 \psi_R^{\ell, j}  \chi _{i} 
   +    h_{ij}^{d}  \bar{\psi}_L^{q,1}   \varphi_2 \psi_R^{d, j}  \chi _1^\prime    \\ \nonumber
   &+&   h_{ij}^{\ell }  \bar{\psi}_L^{\ell, 1}   \varphi_2 \psi_R^{\ell, j}  \chi _1^\prime \Bigr)  
+  {\rm H.c.},
\eea
where $\psi_L^{q,\ell, i}$ are the  SM quark and leptonic doublets with $i=1,2,3$ family indices,  $\psi_R^u =  u_R, c_R, t_R, ~ \psi_R^d = d_R, s_R, b_R$, and  $\psi_R^\ell = e_R, \mu_R, \tau_R $.
\begin{table}[htb]
\begin{center}
\begin{tabular}{|c|c|c|c|}
  \hline
  Fields             &        $\mathcal{Z}_2$                    & $\mathcal{Z}_2^\prime$   & $\mathcal{Z}_2^{\prime \prime}$       \\
  \hline
  $u_{R}, c_{R}, t_{R} $, $  \nu_{e_R}, \nu_{\mu_R}, \nu_{\tau_R}$                 &   -  &     +    & -                    \\
  $d_{R} $, $ s_{R}$, $b_{R}$,  $e_R$, $ \mu_R$,  $\tau_R$  &   +  &     -    & +                \\
  $\chi _1$                        & +  &      -  &+                            \\
   $\chi _1^\prime$                        & -  &      +  &-                              \\
    $\chi _2$                         & -  &      +  &+                           \\
     $\chi _3$                         & +  &      +  &-                          \\
    $ \psi_L^1 $                        & +  &      -  &+                         \\
   $ \psi_L^2 $                         & -  &      +  &+                           \\
   $ \psi_L^3 $                         & + &      +  &-                            \\
   $ \varphi_1 $                         & +  &      -  &+                              \\
   $ \varphi_2 $                         & -  &      +  &-                              \\
  \hline
     \end{tabular}
\end{center}
\caption{The charges of left- and right-handed fermions, Higgs doublets  and gauge singlet scalar fields under $\mathcal{Z}_2$, $\mathcal{Z}_2^\prime$ and $\mathcal{Z}_2^{\prime \prime}$ symmetries.}
 \label{tab_alt}
\end{table} 

The mass matrices of up, down type  quarks and charged leptons can be written as,
\begin{equation}
\label{mu_ap}
\begin{array}{ll}
\M_\U =   \dfrac{ v_2 }{\sqrt{2}}  \left( \begin{array}{ccc}
  y_{11}^u \epsilon_1 &  y_{12}^u \epsilon_1 &   y_{13}^u \epsilon_1 \\
  y_{21}^u \epsilon_2 &   y_{22}^u \epsilon_2 &   y_{23}^u \epsilon_2\\
   y_{31}^u \epsilon_3 &     y_{32}^u \epsilon_3  &   y_{33}^u \epsilon_3\\
\end{array} \right), \\ \nonumber
\M_\D = \dfrac{ 1 }{\sqrt{2}}  \left( \begin{array}{ccc}
( v_1  y_{11}^d +  v_2 h^d_{11} ) \epsilon_1 &    ( v_1  y_{12}^d +  v_2 h^d_{12} ) \epsilon_1 &  ( v_1  y_{13}^d +  v_2 h^d_{13} ) \epsilon_1 \\
 v_1  y_{21}^d \epsilon_2 &  v_1   y_{22}^d \epsilon_2 &  v_1  y_{23}^d \epsilon_2\\
v_1    y_{31}^d \epsilon_3 &  v_1   y_{32}^d \epsilon_3  & v_1  y_{33}^d \epsilon_3\\
\end{array} \right),  \\ \nonumber
\M_\ell =\dfrac{ 1}{\sqrt{2}}  \left( \begin{array}{ccc}
  ( v_1  y_{11}^\ell +  v_2 h^\ell_{11} ) l \epsilon_1 &     ( v_1  y_{12}^\ell +  v_2 h^\ell_{12} ) \epsilon_1 &    ( v_1  y_{13}^\ell +  v_2 h^\ell_{13} ) \epsilon_1 \\
 v_1 y_{21}^\ell \epsilon_2 &  v_1   y_{22}^\ell \epsilon_2 &  v_1  y_{23}^\ell \epsilon_2\\
 v_1    y_{31}^\ell \epsilon_3 &   v_1  y_{32}^\ell \epsilon_3  & v_1  y_{33}^\ell \epsilon_3 \\
\end{array} \right),
\end{array}
\end{equation}
where $\langle \varphi_1 \rangle = v_1$, $\langle \varphi_2 \rangle = v_2$,   $\epsilon_{1,2,3} = \dfrac{\langle \chi _{1,2,3} \rangle }{\Lambda}$,  and $\Lambda$ is an unknown scale.   Moreover, we have assumed $\langle \chi _{1} \rangle = \langle \chi _{1}^\prime \rangle$.  

The masses of charged fermions can approximately be written as,
\begin{eqnarray}
\label{mass_app2}
m_t  &\approx& \ |y^u_{33}| \epsilon_3 v_2/\sqrt{2}, ~
m_c  \approx \   |y^u_{22} - \dfrac{y_{23}^u  y_{32}^u}{y_{33}^u}| \epsilon_2 v_2/\sqrt{2} ,\nonumber \\
m_u  &\approx&  |y_{11}^u - {y_{12}^u y_{21}^u \over m_c}  \epsilon_2 v_2/\sqrt{2} -
{{y_{13}^u (y_{31}^u y_{22}^u - y_{21}^u y_{32}^u )-y_{31}^u  y_{12}^u  y_{23}^u } \over 
{m_c m_t}}  \epsilon_2 \epsilon_3   v_2^2 /2 |\,  \epsilon_1 v_2/\sqrt{2} ,\nonumber \\
m_b  &\approx& \ |y^d_{33}| \epsilon_3 v_1/\sqrt{2}, ~
m_s  \approx \   |y^d_{22} - \dfrac{y_{23}^d  y_{32}^d}{y_{33}^d}| \epsilon_2 v_2/\sqrt{2} ,\nonumber \\
m_d  &\approx&  |( v_1  y_{11}^d +  v_2 h^d_{11} ) \epsilon_1  - {( v_1  y_{12}^d +  v_2 h^d_{12} ) y_{21}^d v_1  \epsilon_1 \epsilon_2 \over m_s \sqrt{2}} \nonumber \\
&-&
{{( v_1  y_{13}^d +  v_2 h^d_{13} )  (y_{31}^d y_{22}^d - y_{21}^d y_{32}^d ) v_1^2 \epsilon_1 \epsilon_2 \epsilon_3  -y_{31}^d  ( v_1  y_{12}^d +  v_2 h^d_{12} )  y_{23}^d v_1^2 \epsilon_1 \epsilon_2 \epsilon_3 } \over 
{2 m_s m_b}}   |\, /\sqrt{2} ,\nonumber \\
m_\tau  &\approx& \ |y^\ell_{33}| \epsilon_3 v_1/\sqrt{2},~
m_\mu  \approx \   |y^\ell_{22} - \dfrac{y_{23}^\ell  y_{32}^\ell}{y_{33}^\ell}| \epsilon_2 v_2/\sqrt{2} ,\nonumber \\
m_e  &\approx&  |( v_1  y_{11}^\ell +  v_2 h^\ell_{11} ) \epsilon_1  - {( v_1  y_{12}^\ell +  v_2 h^\ell_{12} ) y_{21}^\ell v_1  \epsilon_1 \epsilon_2 \over m_\mu \sqrt{2}} \nonumber \\
&-&
{{( v_1  y_{13}^\ell +  v_2 h^\ell_{13} )  (y_{31}^\ell y_{22}^\ell - y_{21}^\ell y_{32}^\ell ) v_1^2 \epsilon_1 \epsilon_2 \epsilon_3  -y_{31}^\ell  ( v_1  y_{12}^\ell +  v_2 h^\ell_{12} )  y_{23}^\ell v_1^2 \epsilon_1 \epsilon_2 \epsilon_3 } \over 
{2 m_\mu m_\tau}}   |\, /\sqrt{2},
\end {eqnarray}
The mass hierarchy among the three fermionic families is achieved when fields $\chi _i$ acquire VEVs in such a way that  $ \langle \chi _{3} \rangle >> \langle \chi _{2} \rangle >> \langle \chi _{1} \rangle $.  The mass hierarchy within the second and third fermionic families  is created by the assuming $\langle \varphi_2 \rangle >> \langle \varphi_1 \rangle$.  The quark mixing angles  approximately are,
\begin{eqnarray}
\sin \theta_{12}  \simeq |V_{us}| &\simeq& \dfrac{\epsilon_1}{\epsilon_2} \left|{ ( v_1  y_{12}^d +  v_2 h^d_{12} )   \over y_{22}^d v_1 }   -{y_{12}^u  \over y_{22}^u }  \right| , 
\sin \theta_{23}  \simeq |V_{cb}| \simeq   \dfrac{\epsilon_2}{\epsilon_3} \left|{y_{23}^d  \over y_{33}^d  }  -{y_{23}^u  \over y_{33}^u  }  \right|,\nonumber \\
\sin \theta_{13}  \simeq |V_{ub}| &\simeq& \dfrac{\epsilon_1}{\epsilon_3}\left|{  ( v_1  y_{13}^d +  v_2 h^d_{13} )  \over v_1 y_{33}^d   }  -{y_{12}^u y_{23}^d   \over y_{22}^u y_{33}^d  } 
- {y_{13}^u  \over y_{33}^u } \right|.
\end{eqnarray}   
\subsection{Neutrino masses and oscillation parameters}
\label{neutrino_app}
The Lagrangian for  Dirac masses for neutrinos  can be written as,
\bea
{\mathcal{L}}_{ \M_{\D} } &=&  \dfrac{1}{\Lambda}       y_{ij}^\nu  \bar{\psi}_L^{\ell,i}  \tilde{\varphi_2} \psi_R^{\nu, j}   \chi _i 
+  {\rm H.c.}, 
\eea
where $\psi_R^\nu = \nu_{e_R}, \nu_{\mu_R}, \nu_{\tau_R} $.  

The Dirac mass matrix  is now  written as,
\begin{equation}
\label{mN}
\begin{array}{ll}
\M_\N = \dfrac{ v }{\sqrt{2}} \left( \begin{array}{ccc}
  y_{11}^\nu \epsilon_1 &    y_{12}^\nu \epsilon_1 &   y_{13}^\nu  \epsilon_1 \\
 y_{21}^\nu \epsilon_2 &    y_{22}^\nu \epsilon_2 &  y_{23}^\nu  \epsilon_2 \\
   y_{31}^\nu \epsilon_3 &   y_{32}^\nu \epsilon_3 &   y_{33}^\nu \epsilon_3
\end{array} \right).
\end{array}
\end{equation}

The  Majorana Lagrangian reads,
\bea
\mathcal{L}_{ \M_{R}}  = M_{ij} \bar{\nu^c}_{i} \nu_{j} 
\eea
The masses of neutrinos  now can be determined using type-\rom{1} seesaw mechanism\cite{seesaw}  providing  following mass matrix of the light neutrinos,
\begin{eqnarray}
{ \M}~ =~ -  \M_{\D}  \M_{R}^{-1}  \M_{\D}^T,
\end{eqnarray}
where $ \M_{\D} << \M_{R}$ is assumed.  The light neutrino masses   approximately are,
\bea
m_3  &\approx& \ |y^\nu_{33}| \epsilon_3 ~ \epsilon^\prime, ~
m_2  \approx \   |y^\nu_{22} - \dfrac{y_{23}^\nu  y_{32}^\nu}{y_{33}^\nu}| \epsilon_2 ~\epsilon^\prime,\nonumber \\
m_1  &\approx&  |y_{11}^\nu - {y_{12}^\nu y_{21}^\nu \over m_2}  \epsilon_2 v_2/\sqrt{2} -
{{y_{13}^\nu (y_{31}^\nu y_{22}^\nu - y_{21}^\nu y_{32}^\nu )-y_{31}^\nu  y_{12}^\nu  y_{23}^\nu } \over 
{m_2 m_3}}  \epsilon_2 \epsilon_3   v_2^2 /2 |\,  \epsilon_1 ~ \epsilon^\prime,
\eea 
where $\epsilon^\prime =v_2/\sqrt{2} M$ and $M$ is a generic Majorana scale.  The leptonic mixing angles are approximately  found to be,
\begin{eqnarray}
\sin \theta_{12}  &\approx& \dfrac{\epsilon_1}{\epsilon_2} \left|{ ( v_1  y_{12}^\ell +  v_2 h^\ell_{12} )   \over y_{22}^\ell v_1 }   -{y_{12}^\nu  \over y_{22}^\nu }  \right| , 
\sin \theta_{23}    \approx   \dfrac{\epsilon_2}{\epsilon_3} \left|{y_{23}^\ell  \over y_{33}^\ell  }  -{y_{23}^\nu  \over y_{33}^\nu  }  \right|, \nonumber \\
\sin \theta_{13}   &\approx& \dfrac{\epsilon_1}{\epsilon_3}\left|{  ( v_1  y_{13}^\ell +  v_2 h^\ell_{13} )  \over v_1 y_{33}^\ell   }  -{y_{12}^\nu y_{23}^\ell   \over y_{22}^\nu y_{33}^\ell  } 
- {y_{13}^\nu  \over y_{33}^\nu } \right|.
\end{eqnarray}   
We note that $\sin \theta_{13} $ is smaller relative to  $\sin \theta_{23} $ and  $\sin \theta_{12} $.  However, it  could be of the order of the Cabibbo angle due to the enhancement coming from the term $v_2/v_1$.  

\subsection{ Benchmark points for fermionic masses, quark-mixing and neutrino oscillation parameters}
The following bench marks points can reproduce masses and mixing of fermions fairly,
\begin{eqnarray}
\{|y_{11}^u|, |y_{12}^u|, |y_{13}^u|,|y_{21}^u|, |y_{22}^u|, |y_{23}^u|,|y_{31}^u|, |y_{32}^u|, |y_{33}^u| \} & = & \{5.5, 12.56, 12.57, 8.92, 0.1, 1.97, 7.65, 0.35, 1.52 \}, \nonumber \\
\{\phi_{11}^u, \phi_{12}^u,  \phi_{13}^u,\phi_{21}^u,\phi_{22}^u,\phi_{23}^u,\phi_{31}^u, \phi_{32}^u,\phi_{33}^u \} & = & \{ 0.76, 5.1, 1, 5.62, 0.21, 5.13, 0.29, 6.23, 0.81\}, \nonumber \\
\{|y_{11}^d|, |y_{12}^d|, |y_{13}^d|, |y_{21}^d|, |y_{22}^d|, |y_{23}^d|, |y_{31}^d|, |y_{32}^d|, |y_{33}^d|\} & = & \{3.86, 8.76, 6.68, 12.17, 0.1, 12.57, 2.92, 4.35, 5.60\}, \nonumber \\
\{\phi_{11}^d, \phi_{12}^d,  \phi_{13}^d,  \phi_{21}^d,\phi_{22}^d,  \phi_{23}^d,\phi_{31}^d \phi_{32}^d,\phi_{33}^d\} & = & \{ 4.39, 1.67, 2.89, 1.98, 0, 0.81, 1.87, 4.28, 5.79    \}, \nonumber \\
\{|y_{11}^\ell|, |y_{12}^\ell|, |y_{13}^\ell|, |y_{21}^\ell|, |y_{22}^\ell|, |y_{23}^\ell|, |y_{31}^\ell|, |y_{32}^\ell|, |y_{33}^\ell|\} & = & \{ 5.62, 12.57, 9.44, 2.95, 0.1, 12.57, 0.1, 0.25, 0.14  \}, \nonumber \\
\{\phi_{11}^\ell, \phi_{12}^\ell,  \phi_{13}^\ell,  \phi_{21}^\ell,\phi_{22}^\ell,\phi_{23}^\ell,\phi_{31}^\ell,\phi_{32}^\ell,\phi_{33}^\ell \} & = & \{ 1.71, 2.86, 1.36, 1.89, 3.18, 1.8, 3.88, 1.98, 5.59  \}, \nonumber \\
\{|h_{11}^\ell|, |h_{12}^\ell|, |h_{13}^\ell |\}  =  \{ 7.9, 12.57, 8.40  \}, &&
\{\delta_{11}^\ell, \delta_{12}^\ell,  \delta_{13}^\ell \}  =  \{ 0.03, 2.81, 1.26  \}, \nonumber \\
\{|h_{11}^d, |h_{12}^d, |h_{13}^d |\}  =  \{ 9.63, 12.57, 2.59  \}, &&
\{\delta_{11}^d, \delta_{12}^d,  \delta_{13}^d \}  =  \{ 5.27, 4.05, 3.71  \}, \nonumber \\
\{|y_{11}^\nu|, |y_{12}^\nu|, |y_{13}^\nu|, |y_{21}^\nu|, |y_{22}^\nu|, |y_{23}^\nu|, |y_{31}^\nu|, |y_{32}^\nu|, |y_{33}^\nu|\} & = & \{1.22, 12.57, 3.07, 4.2, 0.1, 12.57, 7.90, 0.29, 0.49\}, \nonumber \\
\{\phi_{11}^\nu, \phi_{12}^\nu,  \phi_{13}^\nu,\phi_{21}^\nu,\phi_{22}^\nu,\phi_{23}^\nu, \phi_{31}^\nu,  \phi_{32}^\nu,  \phi_{33}^\nu  \} & = & \{4.57, 4.6, 1.86, 4.47, 1.78, 3.78, 4.62, 3.74, 1.29 \}, \nonumber \\
\epsilon_1  =  7.26179 \times 10^{-6},~\epsilon_2  =  6.7452 \times 10^{-3},~\epsilon_3  &=&  0.585162, \delta =  1.196, \nonumber \\
v_1&=& 1 \rm GeV, v_2 = 245.998 GeV,
\end{eqnarray}
where $\delta$ is the Dirac $CP$ phase of the CKM matrix.  

\end{appendix}



\end{document}